\documentclass[twocolumn,showpacs,preprintnumbers,amsmath,amssymb,letterpaper]{revtex4}
\usepackage{amssymb}
\usepackage{graphicx}
\usepackage{dcolumn}
\usepackage{bm}

\begin{document}

\title{Stochastic statistical theory of nucleation and evolution
 of nano-sized precipitates in alloys with application to
 precipitation of copper in iron }
\renewcommand{\abstractname}{}
\author{ K. Yu. Khromov$^a$, F. Soisson$^b$,  A. Yu. Stroev$^a$ and  V. G. Vaks$^a$}
\affiliation{$^a$Russian Research Center "Kurchatov Institute",
123182 Moscow, Russia \\
$^b$Service de Recherches de M\'etallurgie Physique, DMN-SRMP, CEA
Saclay, 91191 Gif-sur-Yvette, France}

\date{\today}

\begin{abstract}

The consistent and computationally efficient stochastic statistical
approach (SSA) is suggested to study kinetics of nucleation and
evolution of nano-sized precipitates in alloys. An important
parameter of the theory is the size of locally equilibrated regions
at the nucleation stage which is estimated using the ``maximum
thermodynamic gain'' principle suggested.
 For several realistic models of iron-copper alloys studied,
the results of the SSA-based simulations of precipitation kinetics
 agree well with the kinetic Monte Carlo simulation results for all main
characteristics of microstructure. The approach developed is also
used to study kinetics of nucleation and changes in microstructural
evolution under variations of temperature or concentration.

\end{abstract}


\maketitle

\section{ INTRODUCTION}

Studies of microstructural evolution in phase-separating  alloys
attract interest from both fundamental and applied points of view.
>From the fundamental side, elucidation of microscopic mechanisms for
the formation and evolution of embryos  of  new phases arising under
the first-order phase transitions is one of the principal problems
in statistical physics being not well understood as yet
\hbox{\cite{LL,Lif-Pit,Binder,SM-00}}. From the applied side,
understanding factors that determine different characteristics of
microstructure formed under precipitation  is important to control
these characteristics, particularly for alloys with nano-sized
precipitates which attract recently much attention in connection
with industrial applications
\cite{Miller-03,Cerezo-05,Isheim-06,Rana-07}.

Presently, theoretical studies of the precipitation kinetics employ
 usually either the  phase-field method (PFM)
 \cite{WBSK-98,Guo-05,Koyama-05,Nagano-06,Zhang-06,Bronchart-08}
 or Monte Carlo modeling \cite{SM-00,SBM-96,LBS-02,Clouet-05,SF-07}.
However, employing the phase-field methods to describe nucleation
and evolution of nano-sized precipitates can be misleading for at
least three reasons. First, the ``continuous'' approximation used in
the PFM disregards the discrete lattice effects which should be
important at first stages of nucleation when typical precipitate
sizes are few lattice constants. Second, at the concentration and
temperature values typical for applications, the mean-field-type
CALPHAD expressions for thermodynamic potentials usually employed in
the PFM studies \cite{WBSK-98,Guo-05,Koyama-05,Nagano-06,Zhang-06}
will be shown to strongly distort the position of spinodals, thus
using these expressions can drastically distort the type of
microstructural evolution. Third, treatment of fluctuative terms
(being crucial for the nucleation stage) in the ``stochastic PFM''
versions used in applications until now \cite{WBSK-98,Guo-05} seems
to be arbitrary and inconsistent \cite{DPSV,SPV-08}, while the
adequate description of these terms determines all main
characteristics of microstructure.

 Therefore, the only reliable source of theoretical information about
nucleation and evolution of nano-sized precipitates now is Monte
Carlo modeling, in particular, the kinetic Monte Carlo approach
(KMCA) developed in Refs.
\cite{SM-00,SBM-96,LBS-02,Clouet-05,SF-07}. However, present
versions of the KMCA are time-consuming, which may partly explain a
relatively few number of applications of this method to concrete
systems \cite{Clouet-05,SF-07}. Moreover, the lattice misfit and
elastic strain effects important for many phase-separating alloys
can not be easily taken into account in the KMCA, while it makes no
problem for statistical approaches \cite{Vaks-04}. Finally, in the
KMCA  it is often difficult to follow the dependence of various
characteristics of evolution on thermodynamic and microscopic
parameters of an alloy, such as the composition, temperature,
different interaction constants, etc, while it is usually much
simpler for statistical methods based on some analytical equations.
Therefore, the development of a consistent statistical theory which
takes into account all achievements of the KMCA seems to be very
important for a more deep understanding of the phase separation
kinetics.

Recently Sroev et al. \cite{SPV-08} (SPV) presented an attempt to
develop such a theory using the stochastic statistical approach
(SSA) described below in Sec. III B. To illustrate the main ideas of
this approach, SPV used only simplest methods and models, such as
the mean-field approximation (MFA),
 continuous approximations, the direct-atomic-exchange rather
than the vacancy exchange kinetic model, oversimplified interaction
models, etc, while no attempts of quantitative treatments for
realistic alloy models have been made.

The main aim of the present work is to raise the accuracy and the
predictive power of the SSA in describing the precipitation kinetics
for realistic alloy models up to the level comparable to that of the
KMCA. To this end, we perform detailed studies of nucleation and
evolution of nano-sized precipitates in Fe-Cu alloys using both the
KMCA \cite{LBS-02,SF-07} and the SSA. This requires many refinements
of simple models used by SPV.  We have to consider the realistic
vacancy-mediated exchange kinetics rather than the simplified
direct-atomic-exchange model; to use the quantitative, cluster
statistical methods rather than the simple MFA; to allow for strong
concentration and temperature dependences of generalized mobilities
in the resulting kinetic equations, etc. All these refinements are
made in the present work. We also introduce the ``maximum
thermodynamic gain'' principle to determine the key kinetic
parameter of the SSA, the characteristic length of local equilibrium
in the course of the nucleation process.

 In Sec. II we discuss the methodological problems mentioned above:
generalizations of our statistical approach to the vacancy-mediated
kinetics case; employing cluster methods for both thermodynamic and
kinetic statistical calculations; elaboration of effective methods
for calculations of effective mobilities which enter resulting
kinetic equations, etc. Here we also generalize the
earlier-suggested ``equivalence theorem'' \cite{BV-98} which enables
us to reduce the vacancy-mediated kinetics to that for some
equivalent direct-exchange models. In Sec. III we remind basic ideas
of the classical theory of nucleation and present the main equations
of the SSA. In Sec. IV we discuss the models and the methods of
simulations used and we describe the ``maximum thermodynamic gain''
principle suggested to estimate the local equilibrium length
mentioned above. In Sec. V we discuss the features of
microstructural evolution observed in the KMCA and SSA  simulations
for various alloy states considered.  Here we also use the SSA  to
study kinetics of nucleation and influence of variations of
temperature or concentration on evolution of microstructure. Our
main conclusions are summarized in Sec. VI.

\section{QUASI-EQUILIBRIUM KINETIC EQUATION FOR VACANCY-MEDIATED
KINETICS IN SUBSTITUTIONAL ALLOY}

\subsection{General equations for mean occupations of lattice sites}

In this section we derive kinetic equations for mean occupations of
lattice sites disregarding fluctuations of atomic fluxes which lead
to local violations of the second law of thermodynamics. These
``quasi-equilibrium kinetic equations'' (QKE) differ from the
stochastic kinetic equations discussed below (which take into
account such fluctuations) and generalize those  used by SPV
\cite{SPV-08} for simplified direct-atomic-exchange models to more
realistic, vacancy-mediated-exchange (VME) models. Here we also
generalize a similar treatment of the VME kinetics made by
Belashchenko and Vaks \cite{BV-98} (BV) to more realistic VME models
used  in Refs. \cite{LBS-02,SF-07} and below.\,

 We consider a  substitutional alloy with $(m+1)$ components
\,$p'$\,  which includes atoms of $m$ different species ${\rm
p}={\rm p}_1,{\rm p}_2,\ldots {\rm p}_m$ and vacancies v: \,$
p'=\{{\rm p},{\rm v}\}$\,. The distributions of atoms on the lattice
sites $i$ are described by the different occupation number sets
$\{n_i^{p'}\}$ where the operator $n_i^{p'}$ is 1 when the site $i$
is occupied by a \,$p'$-species component and 0 otherwise. For each
\,$i$\, these operators obey the identity \,$\sum_{p'}n_i^{p'}=1$,\,
so only \,$m$\, of them are independent. It is convenient to mark
the independent operators with greek letters  \,$\rho$\,  or
\,$\sigma$\,:
 \,$(n_i^{p'})_{\rm indep}=n_i^{\rho}$,\,
 while the rest operator  denoted as $n_i^h$ is expressed
 via $n_i^{\rho}$:
\begin{equation}
n_i^h=1-\sum_{\rho}n_i^{\rho}.\label{n_h}
\end{equation}
For dilute alloys, it is convenient to put ``$h$" in (\ref{n_h}) to
be the host component, e.g., $h$$=$Fe for  FeCuv alloys.

In terms of all operators $n_i^{p'}$ the configurational Hamiltonian
$H'$ (for simplicity supposed to be pairwise) can be written as
\begin{equation}
H'={1\over
2}\sum_{p'q',ij}V_{ij}^{p'q'}n_i^{p'}n_j^{q'}.\label{H-prime}
\end{equation}
After elimination of operators $n_i^h$ according to (\ref{n_h}),
 the Hamiltonian
(\ref{H-prime}) takes the form
\begin{eqnarray}
&& H= E_0+\sum_{\rho i}\varphi_{\rho}n_i^{\rho}+H_{int}\nonumber\\
&& H_{int}=\sum_{\rho\sigma,
i>j}v_{ij}^{\rho\sigma}n_i^{\rho}n_j^{\sigma} \label{H_int}
\end{eqnarray}
which includes only independent $n_i^{\rho}$, while constants $E_0$,
$\varphi_{\rho}$, and ``configurational interactions''
$v_{ij}^{\rho\sigma}$ are linearly expressed via the interactions
$V_{ij}^{p'q'}$ in (\ref{H-prime}), in particular:
\begin{equation}
v_{il}^{\rho\sigma}=(V^{\rho\sigma}-V^{\rho{h}}-V^{{h}\sigma}
+V^{hh})_{ij}.\label{varphi_i}
\end{equation}
 The fundamental master equation for the probability $P$
of finding the occupation number set $\{n_i^{\rho}\}=\xi$ is
\cite{Vaks-04}:
\begin{equation}
dP(\xi)/ dt=\sum_{\eta} [W(\xi,\eta)P(\eta)-
W(\eta,\xi)P(\xi)]\equiv\hat SP\label{dP/dt}
\end{equation}
where \,$W(\xi,\eta)$\, is the $\eta\rightarrow\xi$ transition
probability per unit time. Adopting   for probabilities \,$W$\, the
conventional ``transition state'' model
\cite{SM-00,SBM-96,LBS-02,SF-07}, we can express the transfer matrix
\,$\hat S$\, in  (\ref{dP/dt}) in terms of the probability of an
elementary inter-site exchange (``jump'') p$i\rightleftharpoons {\rm
v}j$\, between neighboring sites \,$i$\,  and \,$j$:\,
\begin{equation}
W_{ij}^{\rm pv}=n_i^{\rm p}n_j^{\rm v}\omega_{\rm
pv}\exp[-\beta(\hat E_{{\rm p}i,{\rm v}j}^{SP}-\hat E_{{\rm p}i,{\rm
v}j}^{in})]\label{W_ij^pv}
\end{equation}
where  \,$\omega_{\rm pv}$ \, is the attempt frequency, \,$\beta
=1/T$\, is the reciprocal temperature,
 \,$\hat E_{{\rm p}i,{\rm v}j}^{SP}$\, is the
saddle point energy, and $\hat E_{{\rm p}i,{\rm v}j}^{in}$ is the
initial  (before the jump) configurational energy of a jumping atom
 \,$\rm p$\, and a vacancy. The saddle point energy \,$\hat E_{{\rm p}i,{\rm v}j}^{SP}$,\,
 generally, depends on the atomic configuration near the bond \,$ij$\, (which
  is neglected in simplified kinetic models \cite{SM-00,SBM-96,BV-98}).
We will describe this dependence by the model used  in Refs.
\cite{LBS-02}  and \cite{SF-07} supposing the saddle-point energy to
depend
 only on occupations of lattice sites \,$l$\, nearest to the center of bond \,$ij$\,
(these sites will be denoted \,$l_{nn}^{ij}$):
\begin{equation}
\hat E_{{\rm p}i,{\rm v}j}^{SP}=\sum_{{\rm
q},\,l=l_{nn}^{ij}}\varepsilon_{\rm q}^{\rm p}\,n_l^{\rm
q}=E_{h}^{\rm p}- \hat\Delta^{\rm p}_{ij}.\label{E_sp}
\end{equation}
Here \,$E_{h}^{\rm p}$\, is the saddle point energy  for the pure
host metal, while the operator\,$\hat\Delta^{\rm p}_{ij}$\,
describes changes in this energy due to a possible presence of
minority atoms near the bond:
\begin{equation}
E_h^{\rm p}=z_{nn}^b\varepsilon_{h}^{\rm p};\qquad \hat\Delta^{\rm
p}_{ij}= \sum_{\rho,\,l=l_{nn}^{ij}}\Delta^{\rm
p}_{\rho}n_l^{\rho}\label{E_h^p}
\end{equation}
where \,$ z_{nn}^b$\, is the total number of nearest lattice sites
 \,$l$\, for each bond (being \,$ z_{nn}^b=6$\, for the BCC lattice), and
\,$\Delta^{\rm p}_{\rho}$\,=\,$(\varepsilon_{h}^{\rm
p}-\varepsilon_{\rho}^{\rm p})$.

As discussed in detail in  \cite{Vaks-04}, for the usual conditions
of phase transitions corresponding to absence of external fluxes of
particles or energy (that is, when the alloy is a ``closed'' but not
an ``open'' statistical system), the  distribution function
\,$P(\xi)=P\{n_i^{\rho}\}$\, in (\ref{dP/dt}) can be written as
\begin{equation}
P\{n_i^{\rho}\}=\exp \Big[\beta(\Omega+\sum_{\rho i}\lambda_i^{\rho
}n_i^{\rho}-H_{int})\Big].\label{P}
\end{equation}
Here parameters \,$\lambda_i^{\rho }$\, (being, generally, both
time- and space-dependent) can be called ``site chemical
potentials'' for \,$\rho$-species atoms; \,$H_{int}$\, is the same
as in (\ref{H_int}); and the generalized grand canonical potential
$\Omega$ is determined by normalization. The analogous equation
(10) of BV \cite{BV-98} (who treated both closed and open systems)
differs from (\ref{P}) with replacing the interaction hamiltonian by
a more general ``quasi-interaction'' operator \,$Q$.\,  For the
closed systems discussed in this work, \,$Q=H_{int}$,\,  and this
 relation greatly simplifies kinetic equations discussed below
with respect to those of BV.

Multiplying Eq. (\ref{dP/dt}) by operators $n_i^{\rho}$ and summing
over  all configurational states, i.e. over all number sets
$\{n_i^{\rho}\}$, we obtain the set of equations for mean
occupations of sites (``local concentrations'')
 \,$c_i^{\rho}=\langle n_i^{\rho} \rangle$:
\begin{equation}
dc_i^{\rho}/dt= \langle n_i^{\rho}\hat S\rangle \label{c_rho-dot}
\end{equation}
where \,$\langle(...)\rangle={\rm Tr}\{(...)P\}$\, means averaging
over the distribution \,$P$,\, for example:
\begin{equation}
c_i^{\rho}=\langle n_i^{\rho}\rangle =
\sum_{\{n_j^{\sigma}\}}c_i^{\rho}P\{n_j^{\sigma}\}. \label{c_rho}
\end{equation}

After a number of manipulations described by BV, Eqs.
(\ref{c_rho-dot}) can be transformed into the QKE for mean
occupations  \,$c_i^{\rho}$.\, These equations are similar to Eqs.
(19) and (42) of BV but include generalizations and simplifications
mentioned above:
\begin{eqnarray}
dc_i^{\alpha}/dt&=&\sum_{j_{nn}(i)}\gamma_{\alpha{\rm
v}}b_{ij}^{\alpha}
(\xi_i^{\rm v}\eta_j^{\alpha}-\xi_j^{\rm v}\eta_i^{\alpha})\label{c_alpha-dot}\\
dc_i^{\rm v}/dt&=&\sum_{j_{nn}(i)}\Big[\xi_j^{\rm v}\Big(\gamma_
{h{\rm v}}b_{ij}^{h}+\sum_{\beta}\gamma_{\beta{\rm v}}b_{ij}^{\beta}\eta_i^{\beta}\Big)\nonumber\\
&& -\{i\to j\}\Big]. \label{c_alpha-v-dot}
\end{eqnarray}
Here and below, greek indices \,$\alpha,\beta\ldots$\, correspond to
the minority atoms;  symbol ``v'' is used for vacancies; index
\,$nn$\, means ``nearest neighbors'', and symbol \,``$j_{nn}(i)$''\,
means summation over sites \,$j$\, being nearest neighbors of site
\,$i$.\,  Term \,$\gamma_ {\rm pv}$\, in (\ref{c_alpha-v-dot})
(where \,p\, is $\alpha$ or $h$, i. e.  a minority or host atom) is
the effective
 exchange rate \,p\,$\rightleftharpoons$\,v\,
 for a pure host metal. This term can be written in the form similar
 to Eq. (\ref{W_ij^pv}):
\begin{equation}
\gamma_ {\rm pv}=\omega_{\rm pv} \exp\,(-\beta E_{ac}^{\rm pv})
\label{gamma^pv}
\end{equation}
where \,$\omega_{\rm pv}$\,  is the same as  in (\ref{W_ij^pv}),
while \,$E_{ac}^{\rm pv}$\, is the effective activation energy which
is expressed via the saddle point energies \,$E_{h}^{\rm p}$\, in
(\ref{E_h^p}) and interactions \,$V_{ij}^{\rm p'q'}$\, in
(\ref{H-prime}) as follows:\,
\begin{equation}
E_{ac}^{\rm pv}=E_{h}^{\rm p}- \sum_j(V_{ij}^{{\rm p}h}+V_{ij}^{{\rm
v}h})+V_{nn}^{hh}. \label{E_ac}
\end{equation}
Note that for minority atoms \,p\,=$\alpha$,\, expressions
(\ref{E_ac}) differ from analogous activation energies \,$E_{ac,{\rm
MC}}^{\rm pv}$\, used in the KMCA and given by Eq. (2.5) in Ref.
\cite{LBS-02}:
\begin{equation}
E_{ac}^{\alpha{\rm v}}=E_{ac,{\rm MC}}^{\alpha{\rm v}}+
v^{\alpha{\rm v}}_{nn};\qquad E_{ac}^{h{\rm v}}=E_{ac,{\rm
MC}}^{h{\rm v}}. \label{E_ac-MC}
\end{equation}
The difference arises because in the statistically averaged Eqs.
(\ref{c_alpha-dot}) and (\ref{c_alpha-v-dot}), the probability
(\ref{W_ij^pv}) is averaged over distribution (\ref{P}), and for the
inter-site exchange  $\alpha i\rightleftharpoons {\rm v}j$,\, it
leads to the additional Gibbs factor
 \,$\exp (-\beta v^{\alpha{\rm v}}_{ij})$\, in the averaged
 probability.

Quantities  \,$b_{ij}^{\rm p}$\, in (\ref{c_alpha-v-dot}) (for
brevity to be called ``correlators'') are certain averages of site
occupations which describe influence of minority atoms in vicinity
of the bond \,$ij$\, on the p$i$$\rightleftharpoons$v$j$  jump
probability:
\begin{equation}
b_{ij}^{\rm p}=\langle n_i^hn_j^h\exp\Big[\sum_{\alpha
l}\beta(u_{il}^{\alpha}+u_{jl}^{\alpha})n_l^{\alpha}+
\sum_{\alpha,\,l=l_{nn}^{ij}}\beta\Delta^{\rm
p}_{\alpha}n_l^{\alpha}\Big]\rangle\label{b_ij^p}
\end{equation}
where \,$\Delta^{\rm p}_{\alpha}$\, is the same as
 in (\ref{E_h^p}), while quantities \,$u_{il}^{\alpha}$\, (to be
called ``kinetic interactions'' as they affect only effective jump
probabilities but not thermodynamic properties) are related to the
interactions \,$V_{ij}^{\rm p'q'}$\, in (\ref{H-prime}) as
follows:\,
\begin{equation}
u_{il}^{\alpha}=V_{il}^{\alpha h}-V_{il}^{hh}. \label{u_ij^alpha}
\end{equation}
Finally, quantities    \,$\xi_i^{\rm v}$\, and \,$\eta_i^{\alpha}$\,
in (\ref{c_alpha-v-dot}) and (\ref {c_alpha-dot}) can be called
``site thermodynamic activities'' for   vacancies and
\,$\alpha$\,-species atoms, respectively,  as they are related to
site chemical potentials \,$\lambda_i^{\rho}$\, in (\ref{P}) as:
\begin{equation}
 \xi_i^{\rm v}=\exp\,(\beta \lambda_i^{\rm v}); \qquad
\eta_i^{\alpha}=\exp\,(\beta \lambda_i^{\alpha}),\label{xi_i-eta_i}
\end{equation}
that is, analogously to the relations between conventional
thermodynamic activities  and chemical potentials.

\subsection{Calculations of site chemical potentials  \,$\lambda_i^{\rho}$\,
and correlators   \,$b_{ij}^{\rm p}$}

To solve QKE (\ref{c_alpha-v-dot}), we need explicit expressions for
site chemical potentials
\,$\lambda_i^{\rho}=\lambda_i^{\rho}(c_j)$\, determined by Eqs.
(\ref{c_rho}), and for correlators \,$b_{ij}^{\rm p}=b_{ij}^{\rm
p}(c_k)$\, determined by Eqs. (\ref{b_ij^p}). To find these
expressions, we should use some approximate method of statistical
physics, such as the MFA or cluster methods \cite{Vaks-04}. As
discussed in detail in \cite{VKh-1,VKh-2} and below, at realistic
values of interactions \,$v_{ij}^{\alpha\beta}$\, that significantly
exceed temperature \,$T$\, (which is typical, in particular, for the
iron-copper alloys under consideration), employing the MFA leads to
great errors in calculations of thermodynamic potentials which
exclude any realistic description. At the same time, the pair
cluster approximation (PCA) usually combines simplicity of
calculations with a rather high accuracy, particularly at low
\,$c$\,  and \,$T$\, under consideration, see, \hbox{e. g.}
\cite{VKh-1,VKh-2,VS-99}. Thus, for site the chemical potentials
\,$\lambda_i^{\rho}$\,  we use their PCA-expressions which for a
binary alloy  ABv with minority atoms A, host atoms B, and  a
realistically small concentration of vacancies: \,$c_i^{\rm v}\ll
1$,\,  are given by Eqs. (39) of BV:
\begin{eqnarray}
&&\lambda_i=T\Big[\ln (c_i/c_i^h)+
\sum_{j\neq i}\ln (1-g_{ij}c_j)\Big] \label{lambda_PCA}\\
&&\lambda_i^{\rm v}=T\Big[\ln (c_i^{\rm v}/c_i^h)- \sum_{j\neq i}
\ln (1+g_{ij}^{\rm v}c_j)\Big]. \label{lambda_A-v}
\end{eqnarray}
Here \,$\lambda_i=\lambda_i^{\rm A}$,\,  \,$c_i=c_i^{\rm A}$,\,
\,$c_i^h=(1-c_i-c_i^{\rm v})\simeq (1-c_i)$,\, while the function
\,$g_{ij}$\, or \,$g_{ij}^{\rm v}$\, is expressed via the Mayer
function \,$f_{ij}=[\,\exp\, (-\beta v_{ij})-1$]\, or \,$f_{ij}^{\rm
v}=[\,\exp\, (-\beta v_{ij}^{\rm vA})-1]$\, for the
 potential \,$v_{ij}\equiv v_{ij}^{\rm AA}$\, or
\,$v_{ij}^{\rm vA}$\, defined in (\ref{varphi_i}),
\begin{eqnarray}
 &&v_{ij}=V_{ij}^{\rm AA}-2V_{ij}^{\rm AB}+V_{ij}^{\rm
BB}\nonumber\\
&&v_{ij}^{\rm vA}=V_{ij}^{\rm vA}-V_{ij}^{\rm Bv}- V_{ij}^{\rm
AB}+V_{ij}^{\rm BB}, \label{v_ij-AA,Av}
\end{eqnarray}
 as follows:
\begin{eqnarray}
\hskip-10mm &&g_{ij}=2f_{ij}/[R_{ij}+1+f_{ij}(c_i+c_j)]\nonumber\\
\hskip-10mm &&g_{ij}^{\rm v}=2f_{ij}^{\rm v}/[R_{ij}+1+f_{ij}(c_i-c_j)] \nonumber\\
\hskip-10mm
&&R_{ij}=\left\{[1+(c_i+c_j)f_{ij}]^2-4c_ic_jf_{ij}(f_{ij}+1)\right\}^{1/2}.
\label{g_ij}
\end{eqnarray}

Let us also present the PCA expression for the free energy \,$F$\,
of a binary alloy \cite{Vaks-04} which is used below in Sec. IV B
for discussions of thermodynamics of precipitation:
\begin{equation}
 F =\sum_iT\ln\,
c_i^h-T\frac{1}{2}\sum_{ij}\ln
(1-g_{ij}c_ic_j)+\sum_i\lambda_ic_i\label{F_PCA}
\end{equation}
where \,$g_{ij}$\, is the same as in (\ref{g_ij}). If temperature
\,$T$,\, much exceeds interactions \,$v_{ij}$: \,$T\gg v_{ij}$,\,
the PCA expressions  (\ref{lambda_PCA}) and  (\ref{F_PCA}) transform
into the MFA ones \cite{Vaks-04}. However, as mentioned, in
situations of practical interest we usually have an opposite
inequality: \,$T\ll v_{ij}$,\, and employing the MFA is misleading.

Let us now discuss calculations of correlators \,$b_{ij}^{\rm p}$\,
in (\ref{b_ij^p}). For simplicity, we first consider the case of
configuration-independent saddle-point energies when differences
\,$\Delta^{\rm p}_{\alpha}$\, in Eqs.
 (\ref{E_h^p}) and  (\ref{b_ij^p}) vanish and correlators
\,$b_{ij}^{\rm p}=b_{ij}$\, do not depend on the kind of a jumping
atom  \,p.\, Using identities
\begin{eqnarray}
\hskip-10mm&&(n_l^{\alpha})^2=n_l^{\alpha},\quad
n_l^{\alpha}n_l^{\beta}
=n_l^{\alpha}\delta_{\alpha\beta},\quad n_l^{\alpha}n_l^h=0,\nonumber\\
\hskip-10mm&&\exp\,(xn_l^{\alpha})=1+n_l^{\alpha}f(x),\quad
f(x)=(e^x-1), \label{f_x}
\end{eqnarray}
we can rewrite Eq. (\ref{b_ij^p}) for this case as follows:
\begin{eqnarray}
\hskip-10mm&b_{ij}&=\Big\langle n_i^hn_j^h\prod_{l=1}^{k_t}
(1+\sum_{\alpha}f_l^{\alpha}n_l^{\alpha})\Big\rangle= \nonumber\\
\hskip-15mm&&\sum_{k=0}^{k_t}\hskip2mm\sum_{l_1\neq\ldots
l_k}\sum_{\alpha_1\ldots\alpha_k}\Big\langle n_i^hn_j^h
n_{l_1}^{\alpha_1}\ldots n_{l_k}^{\alpha_k}\Big\rangle
f_{l_1}^{\alpha_1}\ldots f_{l_k}^{\alpha_k} \label{b_ij-series}
\end{eqnarray}
where quantity \,$f_l^{\alpha}$\, is defined as
\begin{equation}
f_l^{\alpha}=f(\beta u_{il}^{\alpha}+\beta
u_{j\,l}^{\alpha})\label{f_l^alpha}
\end{equation}
with \,$f(x)$\, from (\ref{f_x}),\,  while \,$k_t$\, in
(\ref{b_ij-series}) is the total number of sites with nonzero values
of potentials \,$(u_{il}^{\alpha}+ u_{j\,l}^{\alpha})$.\, For
example,
 for the nearest-neighbor or second-neighbor
interaction models  in the BCC lattice, we have: \,$k_t=14$,\, or:
\,$k_t=20$.\,

To find averages in (\ref{b_ij-series}) over distribution (\ref{P}),
we should again employ some approximate method of calculations, such
as the MFA, PCA or higher-order cluster approximations
\cite{Vaks-04,VS-99}. However,  for the most of systems of practical
interest, in particular, for the iron-copper alloys discussed below,
we can apparently use in (\ref{b_ij-series}) the simple MFA
replacing each operator \,$n_l^{\alpha}$\, by its average value
\,$c_l^{\alpha}=\langle n_l^{\alpha}\rangle$.\, It seems to be
justified as the functions \,$f_l^{\alpha}$\, in Eqs.
(\ref{b_ij-series}) and (\ref{f_l^alpha}) for such systems are
typically rather large (for example, \,$f(\beta u_1)\gtrsim 5$\, for
the systems described by Table III below). Thus the main
contributions to sum (\ref{b_ij-series}) come from averages of
products of many different operators \,$n_l^{\alpha}$\, which
  correspond to well-separated and weakly correlated
  sites \,$l$.\, In particular, for the BCC lattice,  these products
  (even for the nearest-neighbor interaction model)  include terms
with the neighbors from first to tenth, most often third and fourth.
Correlations of occupations for the so distant lattice sites  should
be typically small, thus using the MFA  that neglects such
correlations should be adequate.

In the MFA, Eq.  (\ref{b_ij-series}) takes the form
\begin{equation}
b_{ij}=c_i^hc_j^h\sum_{k=0}^{k_t}\sum_{l_1\neq \ldots l_m\neq
i,j}\sum_{\alpha_1\ldots\alpha_k}c_{l_1}^{\alpha_1}\ldots
c_{l_k}^{\alpha_k} f_{l_1}^{\alpha_1}\ldots f_{l_k}^{\alpha_k}.
\label{b_ij-MFA}
\end{equation}
To further simplify this expression, we can take into account that
the space variations of local concentration \,$c_l^{\alpha}$\,
arising in the course of alloy decomposition  are  typically rather
smooth, particularly at the nucleation stage (for which an adequate
description of kinetic coefficients that include correlators
\,$b_{ij}$ is most important), see, \hbox{e. g.} Figs.
\ref{nucl-1}-\ref{nucl-3} below.
 Thus, the \,$c_l^{\alpha}$\, values for sites \,$l$\,
adjacent to bond \,$ij$\, that enter  Eq. (\ref{b_ij-MFA}) are
usually close to both the \,$c_i^{\alpha}$\, and \,$c_j^{\alpha}$,\,
as well as to their average
\begin{equation}
\bar c_{ij}^{\,\,\alpha}=
(c_i^{\alpha}+c_j^{\alpha})/2.\label{c_ij^alpha}
\end{equation}
Therefore, to avoid unnecessary complications of computations, we
can approximate each  \,$c_l^{\alpha}$\, in Eq. (\ref{b_ij-MFA}) by
the average (\ref{c_ij^alpha}). It enables us to write the
correlator
 \,$b_{ij}$\, (\ref{b_ij-MFA}) in the simple analytic form:
\begin{equation}
b_{ij}=c_i^hc_j^h\prod_{l\neq i,j}^{k_t}(1+\sum_{\alpha}\bar
c_{ij}^{\,\,\alpha}f_l^{\alpha}). \label{b_ij-product}
\end{equation}
For example, for a binary alloy ABv in the BCC lattice described by
the second-neighbor interaction model with two kinetic interaction
constants, \,$u_1$\, and \,$u_2$,\, we obtain:
\begin{eqnarray}
b_{ij}&=&c_i^hc_j^h\,[1+\bar c_{ij}f(\beta u_1+\beta
u_2)]^6\times\nonumber\\
 &&[1+\bar c_{ij}f(\beta u_1)]^8[1+\bar
c_{ij}f(\beta u_2)]^6 \label{b_ij-u_1,2}
\end{eqnarray}
where index \,$\alpha$\,=A\, in \,$c_{ij}^{\,\,\alpha}$\, and
\,$u_n^{\alpha}$\, is omitted for brevity.

When differences \,$\Delta^{\rm p}_{\alpha}$\, in Eqs.
 (\ref{E_h^p}) and  (\ref{b_ij^p}) are nonzero, the correlator
\,$b_{ij}^{\rm p}$\, in Eq. (\ref{b_ij^p}) can be calculated by the
same way as \,$b_{ij}$\, in Eqs.
(\ref{b_ij-series})-(\ref{b_ij-u_1,2}). The difference arises only
for sites \,$l=l_{nn}^{ij}$\, adjacent to bond \,$ij$\, for which
factors \,$f_l^{\alpha}$\, defined by Eq. (\ref{f_l^alpha}) are now
replaced by analogous factors \,$f_{\Delta}^{\alpha {\rm p}}$\,
defined as:
\begin{equation}
f_{\Delta}^{\alpha {\rm p}}=f(\beta u_{il}^{\alpha}+\beta
u_{j\,l}^{\alpha}+\beta\Delta^{\rm p}_{\alpha}).\label{f_l^alpha-p}
\end{equation}
In particular,  for the BCC binary alloy ABv with the
second-neighbor interaction, we obtain instead of
(\ref{b_ij-u_1,2}):
\begin{eqnarray}
&b_{ij}^p=&c_i^hc_j^h\,[1+\bar c_{ij}f_{\Delta}^{\rm Ap}]^6\times\nonumber\\
&&[1+\bar c_{ij}f(\beta u_1)]^8[1+\bar c_{ij}f(\beta u_2)]^6
\label{b_ij^p-delta}
\end{eqnarray}
where  \,$f_{\Delta}^{\rm Ap}$= $f[\beta (u_1+u_2+\Delta^{\rm
p})]$,\, and \,$\Delta^{\rm p}$\,=\,$(\varepsilon_{h}^{\rm
p}-\varepsilon_{\rm A}^{\rm p})$.

\subsection{Equivalence of  precipitation kinetics for the
 vacancy-mediated exchange models to that for certain
direct exchange models}

In this section we show that  the VME kinetics described by Eqs.
(\ref{c_alpha-dot}) and (\ref{c_alpha-v-dot}) can usually be
described in terms of certain equivalent direct-atomic-exchange
(DAE) models. It will generalize the analogous ``equivalence
theorem'' derived by BV.

First  we note that the vacancy activity \,$\xi_i^{\rm v}=\exp
(\beta \lambda_i^{\rm v })$\, in Eqs. (\ref{c_alpha-dot}) and
 (\ref{c_alpha-v-dot}) is proportional to the vacancy
concentration \,$c_i^{\rm v}$.\, It is illustrated by Eqs.
(\ref{lambda_A-v})  and is actually a general relation of
thermodynamics of dilute solutions. Thus time derivatives of mean
occupations are proportional to the local vacancy concentration
\,$c_i^{\rm v}$\, or \,$c_j^{\rm v}$,\, which is natural for the
vacancy-mediated kinetics. As \,$c_i^{\rm v}$\,
 in real alloys is very small, this implies  that the
relaxation times of  atomic distribution \,$\{c_{i}^{\alpha}\}$\,
are by a factor \,${1/c_i^{\rm v}}$\, greater than the time of
relaxation of vacancies
  at the given \,$\{c_{i}^{\alpha}\}$\,
to their ``quasi-equilibrium'' distribution \,$c_i^{\rm v}\{ c_{
i}^{\alpha}\}$\, for which the right-hand side of Eq.
(\ref{c_alpha-v-dot}) vanishes. Therefore, discarding small
corrections of the relative order \,$c_i^{\rm v}\ll 1$,\, we can
rewrite Eq.(\ref{c_alpha-v-dot})
 as follows:
\begin{equation}
0=\sum_{j_{nn}(i)}\Big[\xi_j^{\rm v}\Big(\gamma_{h{\rm
v}}b_{ij}^{h}+\sum_{\alpha}\gamma_{\alpha{\rm
v}}b_{ij}^{\alpha}\eta_i^{\alpha}\Big)-\{i\to j\}\Big].
\label{adiabat}
\end{equation}
which can be called ``the adiabaticity equation'' for the vacancy
activity \,$\xi_i^{\rm v}$.\,  Solving this equation
 we can, in principle, express \,$\xi_i^{\rm v}$\, via
 \,$c_j^{\alpha}$.\, Then substitution of
these  \,$\xi_i^{\rm v}(c_j^{\alpha})$\, into Eq.
\,(\ref{c_alpha-dot})\, yields the QKE for some equivalent DAE
model.

To illustrate this approach, we first consider the VME models with
configuration-independent saddle-point energies. For such models,
parameters \,$\Delta^{\rm p}_{\rho}$\, in (\ref{E_h^p}) are zero,
correlators \,$b_{ij}^{\rm p}=b_{ij}$\,  do not depend on the kind
\,p\, of jumping atom, and the adiabaticity equation (\ref{adiabat})
takes the simple form:
\begin{equation}
\sum_{j_{nn}(i)}b_{ij}\,\xi_i^{\rm v}\xi_j^{\rm v}
\Big[\Big(\gamma_{h{\rm v}}+\sum_{\alpha}\gamma_{\alpha{\rm
v}}\eta_i^{\alpha}\Big)/\xi_i^{\rm v}-\{i\to
j\}\Big]=0\label{BV-adiabaticity}
\end{equation}
If we denote the first term in square brackets
(\ref{BV-adiabaticity}) as $1/\nu_i$, then the difference in these
brackets takes the form \,$\nu_i^{-1}-\nu_j^{-1}$.\, Thus the
solution of Eqs. (\ref{BV-adiabaticity}) is provided by $\nu_i$
being a constant independent of the site number \,$i$\, (though
possibly depending on time, as well  as on temperature and other
external parameters):
\begin{equation}
\nu_i=\xi_i^{\rm v}\Big/\Big(\gamma_{h{\rm
v}}+\sum_{\alpha}\gamma_{\alpha{\rm v}}\eta_i^{\alpha}\Big)=\nu
(t).\label{nu_i}
\end{equation}
 Relation (\ref{nu_i}) determines the above-mentioned
``quasi-equilibrium'' vacancy distribution \,$c_i^{\rm
v}\{c_i^{\alpha}\}$\, which adiabatically fast follows the atomic
distribution \,$\{c_i^{\alpha}\}$.\, Substituting it into Eq.
(\ref{c_alpha-dot}) we obtain the explicit kinetic equation for
atomic distributions \,$\{c_i^{\alpha}\}$\, for which influence of
vacancies  is characterized by a single parameter \,$\nu (t)$ being
a ``spatially self-averaged'' quantity:
\begin{eqnarray}
&dc_i^{\alpha}/dt=&\sum_{j_{nn}(i)}b_{ij}\nu(t)\Big[\gamma_{\alpha{\rm
v}}\gamma_{h{\rm v}}\Big(\eta_j^{\alpha}-\eta_i^{\alpha}\Big)\nonumber\\
&&+\sum_{\beta}\gamma_{\alpha{\rm v}}\gamma_{\beta{\rm
v}}\Big(\eta_j^{\alpha}\eta_i^{\beta}-\eta_i^{\alpha}\eta_j^{\beta}\Big)\Big].
\label{c_alpha-DAE}
\end{eqnarray}
The last term of this equation  (missed in the analogous Eq. (46) of
BV) is present only for many-component alloys with two or more
species of minority atoms. Eqs. (\ref{c_alpha-DAE}) can also be
rewritten in the form   used for DAE models \cite{Vaks-04}:
\begin{eqnarray}
\hskip-8mm&&dc_i^{\alpha}/dt=\sum_{j_{nn}(i)}M_{ij}^{\alpha
h}\,2\sinh
[\beta(\lambda_j^{\alpha}-\lambda_i^{\alpha})/2]\nonumber\\
\hskip-12mm&&+\sum_{j_{nn}(i),\,\beta}M_{ij}^{\alpha\beta}2\sinh
[\beta(\lambda_j^{\alpha}+\lambda_i^{\beta}-\lambda_i^{\alpha}-\lambda_j^{\beta})/2]
\label{c_alpha-beta-sinh}
\end{eqnarray}
where generalized mobilities \,$M_{ij}^{\rm pq}$\, which describe
inter-site exchanges \,$\alpha$$\leftrightharpoons$${\rm p}$\, and
\,$\alpha$$\leftrightharpoons$$\beta$\, are given by the following
expressions:
\begin{eqnarray}
\hskip-12mm&&M_{ij}^{\alpha h}=\gamma_{\alpha{\rm v}}\gamma_{h{\rm
v}}\nu(t)\,b_{ij}\exp\,[\beta(\lambda_i^{\alpha}+\lambda_j^{\alpha})/2]\label{M_ij^alpha-h}\\
\hskip-12mm&&M_{ij}^{\alpha\beta}=\gamma_{\alpha{\rm
v}}\gamma_{\beta{\rm
v}}\nu(t)\,b_{ij}\exp\,[\beta(\lambda_i^{\alpha}+\lambda_j^{\alpha}+
\lambda_i^{\beta}+\lambda_j^{\beta})/2]. \label{M_ij^alpha-beta}
\end{eqnarray}
Comparing these expressions to Eq. (32) of BV for mobilities
\,$M_{ij}^{\rm pq}$\, in an alloy with the nearest-neighbor
direct-exchange rates \,$\gamma_{ij}^{\rm pq}$=$\gamma_{\rm pq}$,\,
we see that Eqs. (\ref{M_ij^alpha-h}) and (\ref{M_ij^alpha-beta})
correspond to a DAE model with the following effective direct
exchange rates:
\begin{equation}
\gamma_{\alpha h}^{\rm eff}= \gamma_{\alpha{\rm v}}\gamma_{h{\rm
v}}\,\nu(t);\qquad \gamma_{\alpha\beta}^{\rm eff}=
\gamma_{\alpha{\rm v}}\gamma_{\beta{\rm
v}}\,\nu(t).\label{gamma_pq^eff}
\end{equation}
Note that  the  effective DAE rates (\ref{gamma_pq^eff})
 are by a factor \,$c_{\rm v}$\, smaller than the
vacancy exchange rates \,$\gamma_{\rm pv}$.

Let us now consider more realistic VME models with the
configuration-dependent saddle-point energies when correlators
\,$b_{ij}^{\rm p}$\, in (\ref{b_ij^p}) for different \,p\, are
different.  For such models, the basic adiabaticity equation
(\ref{adiabat}) for vacancy activities \,$\xi_i^{\rm v}$,\,
generally,  can not be solved analytically, thus either numerical or
some approximate analytical methods should be used. Let us discuss
two such approximate methods employed below. For a binary alloy, we
can rewrite Eq. (\ref{adiabat}) in the form:
\begin{equation}
\sum_{j_{nn}(i)}b_{ij}^h\,\xi_i^{\rm v}\xi_j^{\rm v}\gamma_{h{\rm
v}}\Big[\Big(1+\eta_ir_{ij}\Big)/\xi_i^{\rm v}-
\Big(1+\eta_jr_{ij}\Big)/\xi_j^{\rm v}\Big]=0\label{adiabat-i}
\end{equation}
where  \,$\eta_i$=$\eta_i^{\alpha}$=$\exp\,(\beta\lambda_i)$,\, and
\,$r_{ij}$\, is \,$\gamma_{\alpha{\rm
v}}b_{ij}^{\alpha}/\gamma_{h{\rm v}}b_{ij}^h$.\, Eq.
(\ref{adiabat-i}) can be approximately solved if  products
\,$r_{ij}\eta_i^{\alpha}$\, obey either of two inequalities:
\begin{eqnarray}
&&{\rm (a)}\quad \
\eta_ir_{ij}=\exp(\beta\lambda_i)\gamma_{\alpha{\rm
v}}b_{ij}^{\alpha}/\gamma_{h{\rm v}}b_{ij}^h\ll 1;\nonumber\\
&& {\rm (b)}\quad \
\eta_ir_{ij}=\exp(\beta\lambda_i)\gamma_{\alpha{\rm
v}}b_{ij}^{\alpha}/\gamma_{h{\rm v}}b_{ij}^h\gg 1
.\label{r-inequalities}
\end{eqnarray}

In the case (a), the second terms in round brackets in
(\ref{adiabat-i}) are just small corrections to the first ones. In
the zeroth approximation they can be neglected, thus the zero-order
solution of Eq. (\ref{adiabat-i}) is:
\begin{equation}
\xi_i^{\rm v}({\rm i})=\nu(t)\gamma_{h{\rm v}}\label{xi^v-i}
\end{equation}
where the constant factor  \,$\gamma_{h{\rm v}}$\, is introduced so
that the function \,$\nu(t)$\, is analogous to that in (\ref{nu_i}).
Substituting (\ref{xi^v-i}) into (\ref{c_alpha-dot}), we again
obtain Eqs.  (\ref{c_alpha-DAE}) or (\ref{c_alpha-beta-sinh}) for a
binary alloy:
\begin{equation}
dc_i/dt=\sum_{j_{nn}(i)}M_{ij}\,2\sinh
[\beta(\lambda_j-\lambda_i)/2].
 \label{QKE}
\end{equation}
Here  index \,''$\alpha h$''\, at the effective
mobility\,$M_{ij}^{\alpha h}=M_{ij}$\, is omitted for brevity, the
expression for this mobility is similar to that in Eq.
(\ref{M_ij^alpha-h}):
\begin{equation}
M_{ij}({\rm a})=\gamma_{\alpha h}^{\rm
eff}\,b_{ij}^{\alpha}\exp\,[\beta(\lambda_i+\lambda_j)/2],\label{M_ij-a}
\end{equation}
and \,$\gamma_{\alpha h}^{\rm eff}$\,  is the same as in
(\ref{gamma_pq^eff}).

In the case (b), we can rewrite Eq. (\ref{adiabat-i}) as:
\begin{equation}
\sum_{j_{nn}(i)}b_{ij}^{\alpha}\,\xi_i^{\rm v}\xi_j^{\rm
v}\gamma_{\alpha{\rm v}}\Big[\Big(\eta_i+r_{ij}^{-1}\Big)/\xi_i^{\rm
v}- \Big(\eta_j+r_{ij}^{-1}\Big)/\xi_j^{\rm
v}\Big]=0\label{adiabat-ii}
\end{equation}
where the second terms in round brackets are again small corrections
to the first ones.  Thus the zero-order solution of this equation
can be written as: \,$\xi_i^{{\rm v}(0)}=\nu(t)\gamma_{\alpha{\rm
v}}\eta_i$,\, while  corrections are proportional to
\,$r_{ij}^{-1}$.\, However, taking into account these corrections is
necessary to obtain a non-zero right-hand side of Eq.
(\ref{c_alpha-dot}). In finding these small corrections, we can
employ the approximation of a ``smooth distribution of local
concentrations''\, used to proceed from Eq. \,(\ref{b_ij-MFA})\, to
\,(\ref{b_ij-product}),\, that is, we can suppose the \,$r_{ij}$\,
values for all bonds \,$ij$\, of the given site \,$i$\, to be close
to each other:
\begin{equation}
r_{ij}\simeq r_{ii}\simeq r_{jj} . \label{r_i-alpha_j}
\end{equation}
 Then the solution of Eq.
(\ref{adiabat-ii}) with the first-order corrections is provided by
the relation:
\begin{equation}
\xi_i^{\rm v}=\nu(t)\gamma_{\alpha{\rm
v}}\Big(\eta_i+r_{ij}^{-1}\Big). \label{xi^v-ii}
\end{equation}
Substituting (\ref{xi^v-ii}) into (\ref{c_alpha-dot}) we again
obtain Eq.  (\ref{QKE}) but correlator \,$b_{ij}^{\alpha} $\, in the
effective mobility (\ref{M_ij-a}) is replaced by \,$b_{ij}^h $:
\begin{equation}
M_{ij}({\rm b})=\gamma_{\alpha h}^{\rm
eff}\,b_{ij}^h\exp\,[\beta(\lambda_i+\lambda_j)/2]\label{M_ij-b}.
\end{equation}

Physically, the opportunity to reduce the vacancy-mediated kinetics
to the equivalent direct exchange  kinetics is connected with the
above-mentioned fact that in the course of evolution of an alloy,
the distribution of vacancies  adiabatically fast follows that of
the main components. Therefore, one may suppose that such
equivalence holds not only for simplified models  (\ref{nu_i}) or
(\ref{r-inequalities}), but is actually a general feature of the
vacancy-mediated kinetics, while for more general models,
correlators \,$b_{ij}$,\, \,$b_{ij}^{\alpha} $\, or \,$b_{ij}^h $\,
in Eqs. (\ref{M_ij^alpha-h}), (\ref{M_ij-a}) or (\ref{M_ij-b}) are
probably replaced by some more complex expressions with similar
properties.

Function \,$\nu (t)$\, in Eq. (\ref{gamma_pq^eff}) determines the
rescaling of time between the initial VME model and the equivalent
DAE model (\ref{QKE}). Temporal evolution of this DAE model is
actually described by the ``reduced time'' \,$t_r$\,  related to the
real time \,$t$\, by the following differential or integral
relations:
\begin{eqnarray}
\hskip-8mm &&dt_r=\gamma_{\alpha h}^{\rm eff}dt= \gamma_{\alpha{\rm
v}}\gamma_{h{\rm v}}\nu (t)dt; \ \ t_r=\int_0^t
\gamma_{\alpha h}^{\rm eff}(t')\,dt';\label{t_r}\\
\hskip-6mm &&t=\int_0^{t_r}\tau_{\alpha h}^{\rm eff}(t_r')dt_r'
\label{t-t_r}
\end{eqnarray}
where \,$\tau_{\alpha h}^{\rm eff}=(\gamma_{\alpha h}^{\rm
eff})^{-1}$\, has the meaning of the mean time of an atomic exchange
\,$\alpha\leftrightharpoons h$,\, while the variable \,$t_r$\, has a
meaning  of  an effective number of such atomic exchanges. This
natural physical variable is used below in describing the SSA
simulation results.

To find the ``rescaling function''  \,$\nu (t)$\, in Eqs.
(\ref{nu_i})-(\ref{t-t_r}),  one should, generally, compare the
results of simulation of precipitation based on the DAE model
(\ref{QKE}) to those based on the initial VME model.  BV made such
comparison for some simplified model of spinodal decomposition,
while  below we estimate  \,$\nu (t)$\,  for several realistic
models of Fe-Cu alloys using comparison to the KMCA results. Note
that the problem of rescaling of time between the real physical time
and the time units employed in the simulation method used, e. g.,
number of Monte Carlo steps in the KMCA, persists in all simulations
of VME kinetics (see Sec. IV C below), and it strongly depends, in
particular, on the boundary conditions for vacancies adopted in
simulations. For example, BV used the ``vacancy conservation''
model, while in simulations \cite{LBS-02,SF-07} and below, a
possible creation of vacancies at various lattice defects (grain
boundaries, dislocations, etc) is taken into account. Thus, the form
of the
 function \,$\nu (t)$\, depends also on the kinetic model
used for vacancies.

The results presented in Fig. \ref{t_r-t} below show that temporal
variations of \,$\nu (t)$\, can be rather sharp. These variations
arise due to qualitative changes in the distribution of vacancies
with respect to minority atoms related to the phenomenon of
``vacancy trapping'' at interfaces of precipitates discussed in
detail by BV and by Soisson and Fu \cite{SF-07} (SF). The resulting
excess of vacancies near growing or shrinking interfaces leads to an
acceleration of effective exchange rates \,$\gamma^{\rm eff}$\, with
respect to the incubation stage when the precipitates are absent. It
results in an increase of  \,$\gamma^{\rm eff}(t)$\, after beginning
of nucleation, and when the vacancy trapping effect is strong, this
increase can be very large, which is illustrated by Fig. \ref{t_r-t}
below. At the same time, after the nucleation stage is over, the
degree of this trapping does not change significantly. Therefore,
the function \,$\nu(t)$\, can be expected to be approximately
constant before and after nucleation and to monotonously increase
with \,$t$\, in the course of nucleation, as illustrated by Fig. 3
of BV for \,$\nu(t)$\, in their simplified model.

\section{MAIN EQUATIONS OF STOCHASTIC STATISTICAL APPROACH}

\subsection{Basic ideas of the classical theory of nucleation}

Before to describe the SSA, it is convenient to remind the main
ideas of the classical theory of nucleation (CTN)
\hbox{\cite{LL,Lif-Pit,Binder,SM-00}}. The CTN treats embryos of a
new phase within the original metastable one as sufficiently large
objects which arise due to thermodynamic fluctuations. The simplest
version of the CTN considers the embryo as a homogeneous droplet
characterized by its  radius \,$R$,\, the interface energy
\,$\sigma$,\, and the free energy gain (with respect to the original
metastable phase) per unit volume, \,$\Delta f$.\, The excess free
energy needed to form this embryo is:
\begin{equation}
F(R)=4\pi R^2\sigma -(4\,\pi R^3/3)\,\Delta f\,.\label{F_R}
\end{equation}
One of basic notions of the CTN is the critical embryo that can grow
with no further loss of the free energy and thus with no
fluctuations. For the model (\ref{F_R}), it corresponds to the
maximum of
 \,$F(R)$\, with respect to \,$R$,\, thus the critical radius
\,$R_c$\, and the nucleation barrier \,$F_c=F(R_c)$\,
 are:
\begin{equation}
R_c=2\sigma/\Delta f, \qquad F_c=16\pi\sigma^3/3{\Delta
f}^2,\label{R_c-F_c}
\end{equation}
while the probability of the critical fluctuation needed to create
this embryo is estimated according to the thermodynamic fluctuation
theory \cite{LL}:
\begin{equation}
 W_c\sim \exp (-F_c/T)\sim \exp (-{\rm const}\,\,\sigma^3/T\Delta
 f^2).\label{W_c}
\end{equation}

 Cahn and Hilliard \cite{Cahn-Hilliard} used the Ginzburg-Landau-type
  free energy functional to allow for the diffuse character of
 the interface of the critical embryo, but their approach is
valid only at high \,$T\sim T_c$\, and for large embryos when the
discrete lattice effects can be neglected.  Dobretsov and Vaks
\cite{DV-98a,DV-98b} developed a quantitative approach to calculate
thermodynamics of critical embryos which takes into account the
discrete lattice effects and uses the PCA rather than the simple
MFA. Some results of this approach are used below in Table II and
Fig. \ref{crit-embryos}.

For supercritical embryos with \,$R>R_c$,\, the CTN suggests
fluctuation effects to be insignificant. Therefore,  after
completion of nucleation, the main type of microstructural evolution
is growth of  embryos due to the diffusional flux of minority atoms
from the matrix. Later on,
 the evaporation-condensation (or Lifshits-Slyozov-Wagner - LSW)
  mechanism becomes dominant when the larger precipitates grow
at the expense of smaller ones \cite{Lif-Pit}. Therefore, according
to the CTN, decomposition of metastable solid solutions should
include four well-defined stages \cite{SM-00}: (i) the incubation
stage
 that precedes formation of first critical and supercritical
embryos; (ii) the nucleation stage during which the supercritical
precipitate density  \,$d_s$\, reaches its maximum value;\, (iii)
the growth stage when the density \,$d_s(t)$\, remains approximately
constant but sizes  of precipitates grow, and (iv) the coarsening
stage when the density \,$d_s(t)$\, decreases due to the LSW
evaporation-condensation mechanism.

These CTN ideas were confirmed by KMC simulations of Soisson and
Martin \cite{SM-00} (SM) for some simple alloy model for which
critical sizes and nucleation barriers (estimated in Table II below)
are rather large, see, \hbox{e. g.}, Fig. 1 of SM. However, in this
work we mainly consider more realistic models of Fe-Cu alloys
described in Sec. IV for which  nucleation barriers and critical
sizes are not large. Fluctuation effects in such alloys will be
shown below to be strong and important not only for the nucleation
but also  for the growth stage.

\subsection{Stochastic kinetic equation and filtration of noise}

The quasi-equilibrium kinetic equations  (\ref{c_rho-dot}),
(\ref{c_alpha-beta-sinh})  and  (\ref{QKE}) determine evolution of
mean occupations of sites due to the average atomic fluxes across
each bond. However, these averaged equations can describe only those
kinetic processes in which the total free energy \,$F_{tot}$\,
decreases \cite{Vaks-96,BV-98}, while  the nucleation process should
be accompanied by a fluctuative increase of this \,$F_{tot}$\,
needed to overcome nucleation barriers. Therefore, to describe this
process, we should consider fluctuations of atomic fluxes. The
stochastic statistical approach for taking into account such
fluctuations was suggested by SPV \cite{SPV-08}. In this section we
present main equations of this approach while their refinements are
described below in Sec. IV.

We consider a binary alloy for which QKE has the form (\ref{QKE}).
In the SSA, this QKE is replaced by the stochastic kinetic equation
(SKE) which can be conveniently written in the finite difference
form (for a short time interval \,$\delta t$)\, given by Eq. (19) of
SPV:
\begin{equation}
\delta c_i\equiv c_i(t+\delta t)-c_i(t)=\delta
c_{i}^d+\sum_{j_{nn}(i)}\delta n_{ij}^f.\label{SKE}
\end{equation}
Here \,$c_i$\, is the occupation of site \,$i$\, averaged over some
locally equilibrated vicinity of this site, and the ``diffusional''
term $\delta c_{i}^d$ corresponds to the  average atomic transfer to
site \,$i$\, described by the right-hand side of the QKE
(\ref{QKE}):
\begin{equation}
\delta c_{i}^d\{c_k\}=\sum_{j_{nn}(i)}M_{ij}2\sinh
[\beta(\lambda_j-\lambda_i)/2]\,\delta t.\label{delta n^d}
\end{equation}
The last term \,$\delta n_{ij}^f$\, in the SKE \,(\ref{SKE}) is the
fluctuative atomic transfer through the bond \,$(ij$)\, which is
described by the Langevin-noise-type method: each
 $\delta n_{ij}^f$ is treated as a random quantity with the
 Gaussian probability distribution:
\begin{equation}
W(\delta n^f_{ij})=A_{ij}\exp [(-\delta
n^f_{ij})^2/2D_{ij}]\label{W}
\end{equation}
where \,$A_{ij}$\,  is the normalization constant, and  the
dispersion \,$D_{ij}$\, is the same as that for the actual
fluctuative transfer $\delta n^f_{ij}$.\,  This dispersion is
related to the mobility \,$M_{ij}$\,  and the time interval
\,$\delta t$\, in Eq. (\ref{delta n^d}) by the
``fluctuation-dissipation'' type relation (18) of SPV:
\begin{equation}
D_{ij}=\langle(\delta n_{ij}^f)^2\rangle=2M_{ij}\,\delta t.
\label{D_ij}
\end{equation}

Unlike standard applications of the Langevin-noise method to
mechanical systems, for the non-uniform statistical systems under
consideration Eqs. (\ref{SKE})--(\ref{D_ij}) should be supplemented
by the ``filtration of noise'' procedure that eliminates the
short-wave contributions to fluctuations \,$\delta n_{ij}^f$.\, As
discussed in detail by SPV, these contributions  to Eq. (\ref{QKE})
have been already  included in the diffusional term \,$\delta
c_{i}^d$\, which is obtained by statistical averaging just over
these short-wave fluctuations. It agrees with the fact that all
quantities entering Eq. (\ref{SKE}), including the mean site
occupation \,$c_i$,\, site chemical potentials \,$\lambda_i$,\, and
the diffusional term \,$\delta c_{i}^d$,\,   have a physical meaning
only within some locally equilibrated region called in textbooks ``a
quasi-closed subsystem'' \cite{LL} that contains a sufficiently
large number of atoms. In other words, in our statistical
description implying division of the whole non-uniform
non-equilibrium alloy into locally equilibrated quasi-closed
subsystems, we consider only ``thermodynamic'' fluctuations
\,$\delta n_{ij}^f$\, which   have approximately the same value (for
all bonds \,$ij$\, of the given crystal orientation \,$\alpha$)\,
within each quasi-closed subsystem, while a non-zero fluctuative
contribution to the total \,$\delta c_i$\, in  (\ref{SKE}) arises
only due to a relatively weak non-uniformity of these fluctuations.
Therefore, in the last term of Eq. (\ref{SKE}), the full fluctuative
transfer \,$\delta n_{ij}^f$\, should be replaced by its long-wave
(or ``coarse-grained'') part \,$\delta n_{ij}^{fc}$.\, The latter
can be obtained by introducing a proper cut-off factor \,$F_c({\bf
k})$\, in the Fourier-component \,$\delta n_{f\alpha}({\bf k})$\, of
the full fluctuation \,$\delta n_{ij}^f\equiv\delta
n_{\alpha}^f({\bf R}_{s\alpha})$\, where \,${\bf R}_{s\alpha}$\,
denotes the position of the bond \,$ij$\, center in the appropriate
crystal sublattice \,$\alpha$\, formed  by these centers
\cite{SPV-08}:
\begin{eqnarray}
&&\delta n_{\alpha}^{fc}({\bf R}_{s\alpha})=\sum_{\bf k}\exp (-i{\bf
kR}_{s\alpha})\,\delta n_{\alpha}^f({\bf k})\,F_c({\bf k})\nonumber\\
&&\delta n_{\alpha}^f({\bf k})=\frac{1}{N}\sum_{{\bf
R}_{s\alpha}}\exp (i{\bf kR}_{s\alpha})\,\delta n_{\alpha}^f({\bf
R}_{s\alpha}) \label{delta N-fk}
\end{eqnarray}
where \,$N$\, is the total number of lattice sites (or atoms) in the
crystal. The cut-off factor \,$F_c({\bf k})$\, can be taken in the
simple Gaussian-like form which for the BCC lattice looks as
follows:
\begin{equation}
F_c^{\rm BCC }({\bf k})=\exp\,
[-4g^2(1-\cos\varphi_1\cos\varphi_2\cos\varphi_3)]\label{F_c-BCC}
\end{equation}
where \,$\varphi_{\nu}=k_{\nu}a/2$;\, $k_{\nu}$\, is the vector
\,${\bf k}$\, component along the main crystal axis $\nu$; and $a$\,
is the BCC lattice constant.

At large \,$g^2\gg 1$,\, the expression (\ref{F_c-BCC}) is reduced
to a Gaussian \,$\exp\,(-k^2l^2/2)$\, with \,$l=ga$.\,   Thus, the
reduced length \,$g=l/a$\, characterizes the mean size of locally
equilibrated quasi-closed subsystems. Generally, it is the
characteristic length of uniformity of site chemical potentials
\,$\lambda_i$,\, which for a single-phase state (before nucleation)
coincides with  the uniformity length for local concentrations
\,$c_i$.\, Estimates of this size are discussed below in Sec. IV B.

\section{MODELS AND METHODS OF SIMULATIONS}

\subsection{Alloy models and states used for simulations}

In the most of our simulations we use the  first-principle
microscopic model of  Fe-Cu alloys suggested by SF \cite{SF-07} that
describes well available thermodynamic and kinetic data for such
alloys. This is the second-neighbor interaction model based on $ab$
$initio$ Density-Functional Theory calculations using SIESTA code.
For this model, configurational interactions \,$v_n$\, for the
\,$n$-th neighbors  in  (\ref{v_ij-AA,Av}), kinetic interactions
\,$u_n^{\rm Cu}\equiv u_n$\, in  (\ref{b_ij^p}), saddle-point energy
parameters \,$\Delta^{\rm p}$\, in (\ref{b_ij^p-delta}), activation
energies  \,$E_{ac}^{\rm pv}$\, in (\ref{gamma^pv}) (all in eV) and
attempt frequencies \,$\omega_{\rm pv}$\, in  (\ref{gamma^pv}) (in
sec$^{-1}$) have the following values:
\begin{eqnarray}
&&v_1=-0.121+0.182T,\quad v_2=-0.021+0.091T,\nonumber\\
&&u_1=0.127-0.091T,\qquad u_2=0.044-0.045T,\nonumber\\
&&\Delta^{\rm Cu}=0.05,\qquad \Delta^{\rm Fe}=-0.03\nonumber\\
&&E_{ac}^{\rm Cuv}=0.438,\qquad E_{ac}^{\rm Fev}=0.698,\nonumber\\
&&\omega_{\rm Cuv}=5\cdot 10^{15},\qquad \omega_{\rm Fev}=2\cdot
10^{15}. \label{SF}
\end{eqnarray}
Here the second terms in expressions for \,$v_n$\, and \,$u_n$\,
describe  phenomenologically the influence of anharmonic and
magnetic effects. Our simulations for the SF model were mainly
performed at concentration \,$c=0.0134$\, and three temperatures
\,$T$: 773, 713 and 663 K; these alloy states will be denoted as the
\hbox{SF-1}, \hbox{SF-2} and \hbox{SF-3} state, respectively. In
Tables I and II and in Sec. V we also present some results for the
alloy state SF-4 that corresponds to \,$c=0.0197$\, and \,$T$=773 K.

We also employ two more models suggested by Le Bouar and Soisson
\cite{LBS-02} (LBS). Parameters of these models  have been fitted to
energies of configurations and vacancy migration barriers computed
with the Embedded Atom Method potential by Ludwig et al.
\cite{Ludwig}.  The model \hbox{LBS-1} is the nearest-neighbor
interaction model with the following parameter values (in the same
units as in (\ref{SF})):
\begin{eqnarray}
&{\rm LBS-1:}\quad& v_1=-0.20,\quad u_1=0.155,\nonumber\\
&&\Delta^{\rm Cu}=\Delta^{\rm Fe}=0,\nonumber\\
&&E_{ac}^{\rm Cuv}=0.018,\qquad E_{ac}^{\rm Fev}=0.64,\nonumber\\
&&\omega_{\rm Cuv}=\omega_{\rm Fev}=5\cdot 10^{15}. \label{LBS-1}
\end{eqnarray}
For the model \hbox{LBS-2}, all parameters are the same as in
(\ref{LBS-1}) except for the saddle-point-energy parameter
\,$\Delta^{\rm Fe}$\, which is:
\begin{equation}
\hskip-20mm{\rm LBS-2:}\qquad \Delta^{\rm Fe}=-0.238 \label{LBS-2}
\end{equation}
Our simulations for these two models were performed at \,$c$\,=0.01
and \,$T$=1000 K, and  these alloy states
 will be denoted as  the \hbox{LBS-1} and \hbox{LBS-2} state, respectively.

\begin{figure}
\caption {(color online). Calculated phase diagrams for the SF model
(upper figure) and LBS moel (lower figure) described in the text.
Thick lines (red online) correspond to the pair cluster
approximation (PCA), and thin lines  (green online) correspond to
the mean-field approximation (MFA); solid lines are binodals, and
dashed lines are spinodals. Solid circles and chained line show the
binodal and spinodal calculated using the CALPHAD expression for
free energy of bcc Fe-Cu alloys taken from \cite{Koyama-05}.
Triangles indicate the \,$(c,T)$\, states used for simulations.
\label{ph-diag-SF}}
\end{figure}

Comparison of SF and LBS models discussed in \cite{SF-07} (a more
detailed comparison is given in Ref.\cite{Soisson-09}) shows, that
the SF model yields a lower solution energy of copper in BCC iron
(which leads  to the lower density and larger sizes of nucleating
precipitates), and the larger difference between vacancy formation
energies in pure  BCC iron and pure BCC copper (which leads  to a
stronger vacancy trapping in copper precipitates and to a higher
mobility of precipitates). As discussed by SF, the LBS models
describe Fe-Cu alloys less realistically than the SF  model, but we
will consider these LBS models for methodical reasons, to follow the
influence of variations of interaction constants and saddle-point
energy parameters on the precipitation kinetics.

Binodals and spinodals for the SF and LBS models calculated using
the PCA , MFA and CALPHAD expressions for free energy are presented
in Fig. \ref{ph-diag-SF}
 where we also show the alloy states used in our simulations. These
states are chosen in the metastable region \,$T_s(c)<T<T_b(c)$\, or
\,$c_b(T)<c<c_s(T)$\, (where index \,$s$\, or \,$b$\, corresponds to
the spinodal or binodal) which corresponds to the nucleation and
growth (NG) type
 of alloy decomposition. The degree of supersaturation for each of
 these states can be quantitatively characterized by the reduced supersaturation
  parameter \,$s$\, introduced in  \cite{DV-98b}:
\begin{equation}
s(c,T)=[c-c_b(T)]/[c_s(T)-c_b(T)]. \label{s}
\end{equation}
Values  \,$s<1$\, correspond to the NG, while \,$s>1$,\, to the
spinodal decomposition (SD)  evolution type; for the alloy states
studied, values of \,$s$\, are presented in Table I.

To appreciate accuracy of results presented in  Fig.
\ref{ph-diag-SF} we first note that the PCA fully takes into account
the pairwise correlations of atomic positions and neglects only
many-particle ones \cite{VS-99}. Therefore, in the dilute alloy
limit \,\hbox{\,$c_i\to 0$},\, the PCA expressions for thermodynamic
potentials become exact. In particular, the free energy
(\ref{F_PCA}) in this limit takes the form
\begin{equation}
F_{\rm PCA}\Big|_{c\to 0}=-\frac{1}{2}\sum_{ij}c_ic_jf_{ij},
\label{F-c-to-0}
\end{equation}
which is just the discrete lattice analogue of the  first term in
the virial expansion of the free energy in powers of density (in our
case, in powers of \,$c_i$) for the standard ``weakly non-ideal
gas'' theory \cite{LL}. Thus, at low concentrations under
consideration, we can expect the accuracy of all thermodynamic
results of the PCA, including both binodals and spinodals, to be
 high, and for all thermodynamic calculations
we use the PCA.

\begin{center}

\vbox{\noindent TABLE I. Alloy states used in our simulations.
\vskip3mm
\begin{tabular}{|c|cccc|c|}
\hline
&&&&&\\
Model\hskip2mm &\multicolumn{3}{c}{\hskip10mm SF}&&LBS-1,2\\
\hline
&&&&&\\
$T$, \,K&773&713&663&773&1000\\
&&&&&\\
$c$,\ {\rm at}\%&1.34&1.34&1.34&1.97&1\\
&&&&&\\
 $s$&0.285&\hskip2mm 0.352&\hskip2mm 0.425&\hskip2mm 0.426&0.459\\
\hline
&&&&&\\
\hskip2mm Alloy state\hskip2mm &\hskip2mm SF-1\hskip2mm
 &\hskip2mm SF-2\hskip2mm &\hskip2mm SF-3\hskip2mm&\hskip2mm SF-4\hskip2mm  &\hskip2mm LBS-1,2\hskip2mm \\\hline

\end{tabular}}
\end{center}
\vskip3mm

 Fig. \ref{ph-diag-SF} shows that at relatively low \,$c$\, and
\,$T$\, values used in our simulations, the binodals calculated
using the PCA, MFA or CALPHAD methods do virtually coincide with
each other. At the same time, the CALPHAD or  MFA-calculated
spinodals \,$c_s(T)$\,  exceed those found in the PCA (and thus, the
exact ones)  by about twice. It leads to a strong underestimating of
the reduced supersaturation \,$s(c,T)$,\,
 which, in its turn, should result in drastic distortions of
microstructural evolution. Therefore, as mentioned in Sec. I, using
the phase-field type methods based on CALPHAD or MFA expressions for
thermodynamic potentials can hardly provide an adequate description
of alloy decomposition kinetics at low concentrations and
temperatures under consideration.

 Fig. \ref{crit-embryos} illustrates the structure of ``thermodynamic''
critical embryos for the alloy states studied. This structure was
calculated by the method of Ref. \cite{DV-98a} mentioned in Sec. III
A with the use of the PCA. The figure shows, in particular, that the
sizes of critical embryos in our problem are comparable to the host
(BCC iron) lattice constant: \,$a_{\rm Fe}=0.287$\, nm. In Table II
we present main characteristics of these critical embryos: the
reduced nucleation barrier \,$F_c/T$,\, the mean radius of the
embryo, \,$R_c$,\,  the total excess  of minority (copper) atoms in
the embryo with respect to the initial state, \,$\Delta N_c$,\, and
the total number of minority atoms in the embryo, \,$N_c$.\, The
barrier \,$F_c$\, is calculated according to Eq. (4) in
\cite{DV-98b} \,(where it was denoted as \,$\Delta \Omega_c$),\,
while quantities \,$R_c$,\, \,$\Delta N_c$,\, and \,$N_c$\, are
defined by relations:
\begin{eqnarray}
&&R_c^2=\sum_ir_i^2(c_i-c)/\Delta N_c,\qquad  \Delta N_c=\sum_i(c_i-c),\nonumber\\
&&N_c=\Delta N_c+ cN_s^c.\label{R_C-N_c}
\end{eqnarray}
Here index \,$i$\, numbers the lattice sites, \,$N_s^c$\, is the
total number of sites (atoms) in a precipitate, and the term
\,$cN_s^c$\, describes the  contribution  to \,$N_c$\, of   the
``uniform background'' minority atoms. For comparison, in Fig. 3 and
Table II we present also the analogous results for the SM
\cite{SM-00} alloy state which corresponds to the nearest-neighbor
interaction  \,$v_1<0$,\, \,$T=0.4(-v_1)$,\, and \,$c=0.03$.\,
\begin{figure}
\caption{(color online). Concentration profiles  \,$\Delta
c_i=c_i-c$\, in the critical embryos for the alloy states studied.
Different curves from top to bottom correspond to the state SM
\cite{SM-00}, \hbox{SF-1}, \hbox{SF-2}, \hbox{SF-4}, \hbox{SF-3} and
LBS, respectively.\label{crit-embryos}}
\end{figure}

\begin{center}

\vbox{\noindent TABLE II. Parameters of ``thermodynamic'' critical
embryos for the alloy states considered. \vskip3mm
\begin{tabular}{|c||c|c|c|c|c|}
\hline
&&&&&\\
\hskip3mm Alloy state\hskip2mm& \hskip3mm $s$&\hskip3mm $F_c/T$&
\hskip3mm $R_c$, nm&\hskip3mm $\Delta N_c$\hskip3mm &\hskip3mm $N_c$\hskip5mm  \\
\hline
&&&&&\\
SM \cite{SM-00}& \hskip1mm 0.287&7.48&0.488&32.5&\hskip3mm33.8\\
&&&&&\\
SF-1& \hskip1mm 0.285&4.38&0.417&13.3&\hskip3mm14.3\\
&&&&&\\
SF-2& \hskip1mm 0.352&2.47&0.438&9.7&\hskip3mm10.7\\
&&&&&\\
SF-3& \hskip1mm 0.425&1.36&0.468&7.2&\hskip3mm8.2\\
&&&&&\\
SF-4& \hskip1mm 0.426&1.90&0.475&10.4&\hskip3mm 11.9\\
&&&&&\\
LBS-1,2& \hskip1mm 0.459&0.83&0.497 &5.4&\hskip3mm6.2\\
\hline
\end{tabular}}
\end{center}
\vskip3mm

 Let us now discuss the expressions for effective mobilities \,$M_{ij}$\, in
Eq. (\ref{QKE}). Model \hbox{LBS-1} corresponds to the
configuration-independent saddle-point energies for which
correlators \,$b_{ij}^p$\, in (\ref{b_ij^p}) do not depend on the
kind \,p\, of an atom.  As mentioned in Sec. II C, for this case the
adiabaticity equation (\ref{adiabat}) is solved analytically, and
the effective mobility \,$M_{ij}=M_{ij}^{\alpha h}$\, is given by
Eq. (\ref{M_ij^alpha-h}).

For two other models, SF and \hbox{LBS-2}, the saddle-point energies
depend on configurations. Thus the adiabaticity equation
(\ref{adiabat}) can be approximately solved only if either of
inequalities (\ref{r-inequalities}) is obeyed at \,$c_i$\, values
significant for the kinetic process studied; for brevity we denote
such  \,$c_i$\, as \,$c_s$.\, Let us first consider NG-type
processes for the SF model. The experience of our simulations for
all models considered (illustrated by Figs.
\ref{nucl-1}-\ref{nucl-3} below) shows that the ``significant for
NG'' local concentrations \,$c_s$\, have usually the order
\,$c_s\sim$0.1. To estimate products \,$\eta_ir_{ij}$\, in Eqs.
\,(\ref{r-inequalities}),\, we first consider the case when these
\,$c_s$\, are small, thus both the functions \,$\ln(1-g_{ij}c_j)\sim
\ln(1-f_{ij}c_s)$\, in Eq. \,(\ref{lambda_PCA})\, and
\,$(1+\bar{c_{ij}}f^{\rm Ap})^6\sim \exp[6\ln(1+c_sf^{\rm Ap})]$\,
in Eq. \,(\ref{b_ij^p-delta})  can be expanded in powers of
\,$c_s$\,. Substituting  numerical values of
parameters\,(\ref{SF})\,  into these expansions  we obtain the
following estimates of \,$\xi_{ij}=\eta_ir_{ij}$\, for the
\hbox{SF-1} and \hbox{SF-3} states:
\begin{equation}
\xi_{ij}^{\rm SF-1}\sim  20c_s\exp\, (80 c_s),\quad \xi_{ij}^{\rm
SF-3}\sim  40c_s\exp\, (160 c_s) \label{r-SF}
\end{equation}
while  for the \hbox{SF-2} state, the \,$\xi_{ij}$\, value lies
between these two estimates. For \,$c_s\gtrsim$0.1, all these values
much exceed unity.  Therefore, inequality (b) in
(\ref{r-inequalities}) is obeyed, and the QKE (\ref{QKE}) with the
effective mobility \,$M_{ij}$\, given by Eq. (\ref{M_ij-b}) can be
used for all our SF states.

 To estimate products \,$\xi_{ij}=\eta_ir_{ij}$\, in Eqs.
 \,(\ref{r-inequalities})\, for the \hbox{LBS-2} state, we again
 expand functions \,$\ln(1-g_{ij}c_j)\sim \ln(1-f_{ij}c_s)$\, in
\,(\ref{lambda_PCA})\, and
  \,$(1+\bar{c_{ij}}f^{\rm Ap})^6\sim \exp[6\ln(1+c_sf^{\rm Ap})]$\,
in  \,(\ref{b_ij^p-delta})  in powers of \,$c_s$.\, Substituting
numerical values of parameters from Eqs. \,(\ref{LBS-1}),\, we
obtain in this case:
\begin{equation}
\eta_ir_{ij}^{\rm LBS-2}\sim 1400\, c_s\exp\, (-40 c_s).
\label{r-LBS-2-NG}
\end{equation}
At \,$c_s$\, between 0.01 and 0.1, the right-hand side (rhs) of this
estimate is large, inequality (b) in (\ref{r-inequalities}) is
obeyed, and Eq. (\ref{M_ij-b}) for the effective mobility
\,$M_{ij}$\, can be used;  at \,$c_s$$\gtrsim$0,15,\, this rhs is
small, and Eq. (\ref{M_ij-a}) for  \,$M_{ij}$\,  can be used; and at
\,$c_s$\, between 0.1 and 0.15, the rhs is of the order of unity,
thus neither of inequalities (\ref{r-inequalities}) is obeyed.
Therefore, the equivalence of the VME kinetics to that for the DAE
model (\ref{QKE}) can not be formally proved for the LBS-2 state,
and we made no the SSA simulations for this state. However, the
above-mentioned remarks enable us to suggest that actually such
equivalence probably exists, with the mobility \,$M_{ij}$\, in the
QKE (\ref{QKE}) smoothly varying between expression (\ref{M_ij-b})
for \,$c_s$\,$<$0.1 and expression (\ref{M_ij-a})  for
\,$c_s$\,$>$0.15. Then the precipitation kinetics for the LBS-2
state should be quite similar to that for the LBS-1 state differing
only by replacing in Eq. (\ref{M_ij-a}) the factor
\,$b_{ij}^{\alpha}=b_{ij}$\, for the
 LBS-1 state by some smaller factor varying between
 \,$b_{ij}^h$$<$$b_{ij}$\,
at  \,$c_s$\,$<0$.1  and \,$b_{ij}^{\alpha}=b_{ij}$\, at
\,$c_s$\,$>$0.15. It should correspond just to some slowing down of
precipitation kinetics for the LBS-2 state with respect to the LBS-1
state, and the KMCA results of LBS \cite{LBS-02} seem to agree with
these considerations.

To characterize strength of configurational (or ``thermodynamic'')
interactions  \,$v_n$\, and kinetic interactions  \,$u_n$\, for the
alloy states considered, in Table
 III we present values of Mayer functions \,$f_n^v$\,
and of analogous ``kinetic'' functions \,$f_n^u$\, defined by
relations:
\begin{equation}
f_n^v=\exp\,(-\beta v_n)-1,\qquad f_n^u=\exp\,(\beta u_n)-1,
\label{f_n^v-u}
\end{equation}
as well as functions \,$f_{\Delta}^p$\, in Eq. (\ref{b_ij^p-delta}).
Functions \,$f_n^v$\,  enter Eqs. (\ref{lambda_PCA})-(\ref{F_PCA})
for site chemical potentials \,$\lambda_i$\, and the free energy
\,$F$,\, while functions \,$f_n^u$\,  and \,$f_{\Delta}^p$\, enter
Eqs. (\ref{f_x})-(\ref{b_ij^p-delta}),
 (\ref{M_ij^alpha-h}), (\ref{M_ij-a}) and \ref{M_ij-b})  for correlators
 \,$b_{ij}^p$\, and effective mobilities \,$M_{ij}$.\,
In Table III we  also present expressions for correlators
\,$b_{ij}^{\rm Fe}$\, and for reduced effective mobilities
\,$M_{ij}^r= M_{ij}/\gamma^{\rm eff}_{\alpha h}$\,
 in Eqs. (\ref{M_ij^alpha-h}), (\ref{M_ij-a})  and (\ref{M_ij-b}) at
 small \,$c_i\simeq \bar{c}_{ij}\lesssim 1/f_1^v$\,
which have been mentioned to be most significant for the NG-type
processes.

\begin{widetext}
\begin{center}
\vskip3mm \vbox{\noindent TABLE III. Values of functions
\,$f_n^v$,\, \,$f_n^u$,\, \,$f_{\Delta}^p$\, in Eqs.
\protect{(\ref{f_n^v-u})} and \protect{(\ref{b_ij^p-delta})} and
expressions for correlators  \,$b_{ij}^{\rm Fe}$\, and reduced
effective mobilities \,$M_{ij}^r=M_{ij}/\gamma^{\rm eff}_{\alpha
h}$\, in Eqs. (\ref{M_ij^alpha-h}), (\ref{M_ij-a})  and
(\ref{M_ij-b}) at small \,$c_i\simeq \bar{c}_{ij}\lesssim 1/f_1^v$\,
for the alloy states studied. \vskip3mm
\begin{tabular}{|c||c|c|c|c|c|c||c|c|}
\hline
&&&&&&&&\\
\hskip4mm Alloy state\hskip4mm &\hskip4mm $f_1^v$\hskip5mm
&\hskip4mm $f_2^v$\hskip5mm &\hskip4mm $f_1^u$\hskip5mm &\hskip4mm
$f_2^u$\hskip5mm &\hskip4mm $f_{\Delta}^{\rm Fe}$
\hskip5mm &$f_{\Delta}^{\rm Cu}$\hskip5mm &$b_{ij}^{\rm Fe}$&$M_{ij}^r$\\
\hline
&&&&&&&&\\
SF-1&4.1&0.3&5.2&0.9&5.8&24&\hskip2mm $\exp\,(81c_i)$\hskip3mm &\hskip2mm $\exp\,(47c_i)$\hskip3mm \\
&&&&&&&&\\
SF-2&4.9&0.3&6.3&1.0&7.0&33&\hskip2mm $\exp\,(98c_i)$&\hskip2mm $\exp\,(57c_i)$\\
&&&&&&&&\\
SF-3&5.9&0.3&7.5&1.1&8.5&43&\hskip2mm $\exp\,(118c_i)$\hskip3mm &\hskip2mm $\exp\,(69c_i)$\hskip3mm \\
\hline
&&&&&&&&\\
LBS-1&9.2&0&5.0&0&5.0&5.0&$\hskip2mm \exp\,(71c_i)$&\hskip2mm $\exp\,(-3c_i)$\\
\hline
\end{tabular}}
\end{center}
\vskip3mm
\end{widetext}

Let us discuss the results presented in Table III. First, they show
that both the thermodynamic and kinetic interactions for the alloy
systems considered are rather strong: the \,$f_1^v$,\, \,$f_1^u$\,
and \,$f_{\Delta}^p$\, values much exceed unity. It again shows that
the MFA or CALPHAD-type expressions for thermodynamic and kinetic
parameters based on the approximations \,$\beta v_n\ll 1$,\,
\,$\beta u_n\ll 1$\, \cite{Vaks-04} can not be used to describe
these  alloy states. Second, the last column of Table III shows that
the NG kinetics for the SF model  should notably differ from that
for the less realistic  \hbox{LBS-1} model. At small local
concentrations \,$c_i$\, considered, the reduced mobility
\,$M_{ij}^r$\, for the \hbox{LBS-1} model is virtually a constant
close to  unity, while for the SF model it  sharply rises with
\,$c_i$\,  and is typically very large. According to Eqs. (\ref{W})
and (\ref{D_ij}), this mobility
 determines scale of fluctuative terms in the SKE
(\ref{SKE}). Therefore, we can expect the manifestations of
fluctuation effects for the SF model to be  much stronger than for
the \hbox{LBS-1} model. It agrees with the KMCA  results presented
below in Figs. \hbox{\ref{d_p-SF1}-\ref{d_p-LBS1}}.

\subsection{Estimating the local equilibrium length
for the SSA}

As discussed in Sec. III B, the reduced length \,$l=ga$\,
 in the SSA equations (\ref{delta N-fk}) and (\ref{F_c-BCC})
characterizes sizes of locally equilibrium quasi-closed subsystems
used in our statistical description of a nonequilibrium alloy. This
length can not be chosen lower than the characteristic length  of
non-uniformity of local chemical potentials, \,$l_{\rm nu}$,\, which
for the nucleation processes has typically the same order of
magnitude as the critical embryo size \,$R_c$,\, see, \hbox{e. g.},
Figs.  \ref{nucl-1}-\ref{nucl-3} below. The actual distribution of
local equilibrium  lengths \,$l\lesssim l_{\rm nu}$\,  in an alloy
varies with both space and time; in particular, after creation of a
supercritical precipitate, the degree of local equilibrium in the
adjacent region should significantly increase with respect to other
regions where such precipitates are not born yet.

For simplicity we will characterize the distribution  of all local
lengths \,$l$\,  by a  single  spatially averaged parameter \,$\bar
l=ga$\, where the reduced length  \,$g$,\, generally, varies with
the evolution time  \,$t$\, or the reduced time  \,$t_r$\,
(\ref{t_r}) used in the SSA.  After completion of nucleation at some
 \,$t_r=t_r^N$,\, the alloy rapidly approaches
the full two-phase  equilibrium,  and the length  \,$\bar l$\,
should become large. Then   the cut-off  parameter \,$g=g(t_r)$\, in
Eq. (\ref{F_c-BCC})  at  \,$t_r\gtrsim t_r^N$\, should be large,
too, fluctuative terms \,$\delta n_f$\, in  (\ref{SKE}) become
small, and the SKE (\ref{SKE}) transforms into the
 QKE (\ref{QKE}) with no fluctuation terms.

To describe the above-discussed physical picture with the minimal
number of model parameters, we approximated the time dependence
\,$g(t_r)$\, by the following  simplest expression:
\begin{eqnarray}
&& t_r<t_r^N:\quad  g(t_r)=g_0\nonumber\\
&&t_r>t_r^N:\quad   g^2(t_r)=g_0^2+(t_r-t_r^N)CD^{\rm
eff}.\label{g_t-t_r^N}
\end{eqnarray}
Here   \,$D^{\rm eff}$\, is the effective reduced diffusivity which
for the direct-exchange model (\ref{QKE}) can be estimated as
\,$D^{\rm eff}\simeq \gamma^{\rm eff}_{\rm FeCu}$\,
\cite{VBD,BPSV-02}; and \,$C$\, is a numerical factor, \hbox{e. g.},
\,$C$\,=2 for the standard diffusion law. At \,$C\sim 1$,\, the
evolution of microstructure was found to virtually not depend on the
\,$C$\, value, and usually we put \,$C$\,=2.

The simple model (\ref{g_t-t_r^N}) for  \,$g(t_r)$\, includes an
unphysical  break at  \,$t_r=t_r^N$\, which leads to the presence of
analogous fictitious breaks in various characteristics of evolution,
\hbox{e. g.}, in Figs. \ref{F_tg-SF1}-\ref{F_tg-LBS}, \ref{J_t},
\ref{t_r-t} and \ref{i(t)}. As mentioned, in reality the effective
equilibrium length starts to increase  immediately after beginning
of nucleation, thus the function
 \,$g(t_r)=\bar l/a$\, monotonously increases  with time with no breaks. Therefore,
for a more realistic description, we should use  for  \,$g(t_r)$\,
some other  models which describe its continuous increase with
\,$t_r$\, and the resulting smooth decrease of fluctuations in the
course of both nucleation and growth stages. However, the experience
of our simulations shows that at  \,$t_r$\, not close to
\,$t_r^N$,\, the main characteristics of microstructure, such as the
density and sizes of supercritical precipitates, are not sensitive
to the detailed form of \,$g(t_r)$\, provided the scale of this
function is determined by the ``maximum thermodynamic gain''
principle described below. Thus,  in this work we employ for
\,$g(t_r)$\, the  simplest form  (\ref{g_t-t_r^N}).

\begin{figure}
\caption {Total number of precipitates that contain \,$i\geq p$\,
copper atoms, \protect{\,$N_{p}(t_r,g)$},\, versus the reduced time
\,$t_r$\, defined by
 Eq. \protect{(\ref{t_r})} obtained in SSA simulations  with different \,$g$\,
for the \hbox{SF-1} state.\, Frames (a), (b) and (c) correspond to
\,$g$\, equal to 1.75, 1.65 and 1.55, while  different curves from
left to right in each frame correspond to \,$p$\,=10, 20, 40, 60 and
70, respectively.\label{N_p-t}}
\end{figure}

The parameter  \,$g_0$\,  in Eq. (\ref{g_t-t_r^N}) can be estimated
by two ways. First, it can be found by fitting the SSA simulation
results for the evolution of density of precipitates  to the
analogous KMCA results, for example, to those presented in Figs.
\ref{d_p-SF1}-\ref{d_p-LBS1} below. However, such ``KMCA-based''
estimates of   \,$g_0$\,  would restrict possible applications of
the SSA by the models for which reliable KMCA results are available.
To estimate  \,$g_0$\, within the SSA, we can try to extend the
second law of thermodynamics, that is, the principle of the free
energy minimum with respect to all its free parameters valid for
equilibrium systems, to the kinetic processes in non-equilibrium
systems studied. To this end we note that the main characteristics
of microstructure formed in the course of the nucleation process,
such as the characteristic non-uniformity length \,$l_{nu}\sim\bar
l$\, mentioned above, can be considered as ``free'' parameters of a
nonequilibrium state analogous to ``static'' free parameters for
equilibrium systems. Thus it seems natural to suggest that the
kinetic path of evolution of this nonequilibrium state should
correspond to the maximum thermodynamic gain, that is, to the
maximum rate of decrease of free energy. This suggestion can extend
the ``excess entropy production'' approach to thermodynamics of
irreversible processes discussed by Prigogine and coworkers
\cite{Prigogine} to the kinetics of essentially non-uniform and
non-equilibrium systems  under consideration. Then the
characteristic value \,$l_{nu}\sim g_0a$\, can be estimated from the
condition of the maximum thermodynamic gain in the course of the
nucleation process, which in our model \,(\ref{g_t-t_r^N})\,
corresponds to the minimum of the free energy \,$F(g_0,t_r\lesssim
t_r^N)$\, with respect to \,$g_0$.\,

This maximum thermodynamic gain should evidently correspond to the
formation of maximum number of large supercritical precipitates. At
the same time, the density  \,$d_p$\, of such  precipitates sharply
depends on the   effective size \,$\bar l$\, of quasi-equilibrium
systems for the nucleation process. When  \,$g=\bar l/a$\, is too
large, the fluctuations are too weak to overcome nucleation
barriers, while when \,$g$\, is too small, the fluctuations are too
strong to allow formation of large and steadily growing
precipitates. Therefore, the dependences \,$d_p(g)$\,  should have a
pronounced maximum at some optimal \,$g=g_0$.\, These considerations
are illustrated by Fig. \ref{N_p-t} where we present temporal
dependences of the total number of precipitates containing \,$i\geq
p$\, copper atoms, \protect{\,$N_{p}(t_r,g)$},\,  obtained in the
SSA simulations  with different \,$g$.\,
 The ``end of nucleation'' time \,$t_r^N$\, here
and below is defined as the time of creation of the last
``critical'' precipitate with \,$p\geq p_c$,\, while these \,$p_c$\,
values (which for SF alloy states are close to the $N_c$ values in
Table II) are estimated in Sec. V A. Such definition of \,$t_r^N$\,
is somewhat arbitrary but it makes virtually no effect on the
simulation  results. For the evolution shown in frame \ref{N_p-t}a,
the fluctuations seem to be too weak, thus the length \,$\bar
l=ga$\, is too large. For frame  \ref{N_p-t}c, the fluctuations are
evidently too strong which leads to an unphysical evolution of large
precipitates: they do not grow, in disagreement with the second law
of thermodynamics; thus, the \,$g$\, value here is too small.
Finally, for the evolution
 shown in frame  \ref{N_p-t}b, the \,$g$\, value seems to be close to optimum.

\begin{figure}
\caption {(color online). Temporal dependence of the free energy per
copper atom, \,$F(t_r,g_0)$,\, obtained in SSA simulations  with
different \,$g_0$\, for the \hbox{SF-1} state. Curves 1, 2, 3. 4 and
5 (red, green, blue, purple and black online) correspond to
\,$g_0$\,=1.55, 1.65, 1.7, 1.75 and 1.85,
respectively.\label{F_tg-SF1}}
\end{figure}
\begin{figure}
\caption {(color online). Same as in Fig. \protect{\ref{F_tg-SF1}}
but for the SF-2 state.
 Curves 1, 2, 3 and 4  (red, green,  blue and purple online)
correspond to  \,$g_0$\,=1.6, 1.7, 1.8 and 1.9,
respectively.\label{F_tg-SF2}}
\end{figure}
\begin{figure}
\caption {(color online). Same as in Fig. \protect{\ref{F_tg-SF1}}
but for the SF-3 state.
 Curves 1, 2, 3 and 4 (red, green,  blue  and purple online)
correspond to  \,$g_0$\,=1.8, 1.9, 2 and 2.1,
respectively.\label{F_tg-SF3}}
\end{figure}

\begin{figure}
\caption {(color online). Same as in Fig. \protect{\ref{F_tg-SF1}}
but for the LBS-1 state.
 Curves 1, 2, 3. 4 and 5 (red, green, purple, blue and black online)
correspond to  \,$g_0$\,=1.7, 1.8, 1.9, 2 and 2.1,
respectively.\label{F_tg-LBS}}
\end{figure}

In Figs. \ref{F_tg-SF1}-\ref{F_tg-LBS} we present temporal
dependences of the free energy per copper atom, \,$F(t_r,g_0)$,\,
obtained in the SSA simulations with different \,$g_0$\, in Eq.
(\ref{g_t-t_r^N}). For deiniteness, the initial state for these
simulations was taken uniform: \,$c_i(0)=c$\,=const, thus the
initial increase of \,$F$\, at \,$t_r\lesssim 0.1\,t_r^N$\, is
related just to switching-on fluctuations at \,$t_r=0$\, and has no
physical meaning. At \,$t_r\simeq t_r^N$,\, functions
\,$F(t_r,g_0)$\, show the above-mentioned fictitious breaks due the
analogous breaks in our model function \,$g(t_r)$\, in
(\ref{g_t-t_r^N}). However, at \,$t_r$\, not close to zero or to
\,$t_r^N$,\, say at \,$t_r\sim 0.5\,t_r^N$,\,
 functions \,$F(t_r,g_0)$,\, presented in Figs. \ref{F_tg-SF1}-\ref{F_tg-LBS}
can be realistic. Therefore, the ``optimal'' \,$g_0$\, value can be
estimated from comparison of these functions at \,$t_r\sim
0.5t_r^N$\, for different \,$g_0$.\,

 Figs. \ref{F_tg-SF1}-\ref{F_tg-LBS} show that these functions
have a distinct minimum at certain  \,$g_0$\, for each alloy state
considered. For the SF-1 state, this minimum  corresponds to
\,$g_0$\, equal to 1.65\, or 1.7;\, for the SF-2 state,  to
\,$g_0$\,=1.7\, or \,1.8;\, for the SF-3 and LBS-1 states, to
\,$g_0$\,= 1.9\, or 1.8. The first values seem to be a bit more
appropriate, but using in simulations the second values changes
results only slightly; it is illustrated below for the SF-2 model.
Therefore, for the reduced length  \,$g_0$\, in Eq.
(\ref{g_t-t_r^N}) we use the following values:
\begin{eqnarray}
&&g_0^{\rm SF-1}=1.65;\quad  g_0^{\rm SF-2}=1.7;\,\nonumber\\
&&g_0^{\rm SF-3}=1.9;\quad  g_0^{\rm LBS-1}=1.9.\label{g_0-numbers}
\end{eqnarray}
The minimum local equilibrium lengths  \,$l_0=g_0a$\, for these
 \,$g_0$\, have the same order of magnitude as critical sizes  \,$R_c$\,
 in Table II, in accordance with  the considerations mentioned above.

Let us comment on the loss of validity of the SSA at low
 \,$g$\, which is manifested, in particular, in the above-mentioned
unphysical results presented in frame \ref{N_p-t}c. This problem was
discussed in detail by SPV \cite{SPV-08} who noted that the
statistical approach used, in particular, basic equations
(\ref{P})-(\ref{c_alpha-v-dot}) that include averaging over locally
equilibrated quasi-closed subsystems, imply their reduced
 size  \,$g$\, to be not too small, so that they
include a sufficiently large number of atoms, while site chemical
potentials \,$\lambda_i$\,
  within these subsystems should obey the local equilibrium condition
\,$\lambda_i \simeq$ const. The scale of violations of this basic
condition can be characterized by the ``parameter of
non-equilibrium'' \,$J$\, introduced by SPV:
\begin{equation}
J(g,t_r)=\frac{1}{N_{b}}\sum_{i,j}|\lambda_i-\lambda_j|/T
\label{J_g-t}
\end{equation}
where \,$N_b$\, is the total number of the nearest-neighbor bonds
\,$\{ij\}$, and the sum is taken over all such bonds in an alloy. In
Fig. \ref{J_inc-g} we show the values of this parameter  averaged
over incubation stage, \,$J_{\rm inc}(g)=\langle J\,\rangle_{\rm
inc}$;\, for different alloy states, functions \,$J_{\rm inc}(g)$\,
are similar. At small \,$g\lesssim 1.5$,\, these functions start to
sharply rise which  reflects sharp violations of statistical
equilibrium within too small quasi-closed subsystems. However, for
\,$g$=$g_0$\, values presented in Eqs. (\ref{g_0-numbers}), this
parameter is still small: \,$J_{\rm inc}(g)\sim \hbox{0,1-0.15}$,\,
thus employing the SSA seems to be justified.

\begin{figure}
\caption {(color online). Parameter of non-equilibrium
\,\protect{(\ref{J_g-t})}\, for the incubation stage  observed in
SSA simulations with different reduced lengths \,$g$.\, Circles and
squares (red and black online)
 correspond to the \hbox{SF-1} and \hbox{LBS-1} state, respectively.
 Arrows indicate \,$g_0$\, values in Eqs. \,\protect{(\ref{g_0-numbers})}.
 \label{J_inc-g}}
\end{figure}
\begin{figure}
\caption {(color online). Parameter  \,$J(g,t_r)$\,
\,\protect{(\ref{J_g-t})}\,
 observed in the SSA simulations with \,$g$=$g_0$\, from
\,\protect{(\ref{g_0-numbers})}.\, Curves 1, 2, 3 and 4 (red, green,
blue and purple online) correspond to the SF-1, SF-2, SF-3 and LBS-1
alloy state, respectively. \label{J_t}}
\end{figure}

In Fig. \ref{J_t} we show temporal dependences of  the
nonequilibrium parameter (\ref{J_g-t})  at first stages of
evolution. As mentioned, breaks in curves \,$J(t_r)$\, at
\,$t_r=t_r^N$\, are due to the similar breaks in our simple model
(\ref{g_t-t_r^N}), while for more realistic models mentioned above
these breaks should be replaced by some smooth decrease of
\,$J(t_r)$\, at  \,$t_r\gtrsim 0.5\,t_r^N$.\,
 The increase of \,$J(t_r)$\, after beginning of nucleation
is related to arising of interfaces and spacial non-uniformities
which lead to some additional, ``non-uniform'' contributions to
differences \,$|\lambda_i-\lambda_j|$\, in Eq. (\ref{J_g-t}).

\subsection{Methods of KMCA simulations}

The KMCA used in this study is described in detail in Refs.
\cite{LBS-02,SF-07}. We just recall here the physical principles of
the method, underlining the difference with the SSA. The KMC
simulations follow the evolution of the atomic configuration in a
simulation box of
 \,N = $64^3$\,  lattice sites containing Fe and Cu atoms and one
vacancy, with periodic boundary conditions. At each Monte Carlo
step, 8 atom-vacancy exchanges between nearest-neighbor sites can
occur in a BCC lattice, with the jump frequencies  $W_{ij}^{pv}$
given by Eq. (\ref{W_ij^pv}). The activation energies are exactly
computed for each local configuration, without using any mean-field
approximation. One of these exchanges is chosen, using a
pseudo-random generator, by means of a residence time algorithm
\cite{Young-66}. The physical time of the Monte Carlo simulation is
given by:
\begin{equation}
t_{MCS}= 1\Big/\sum_{j=1}^8 W_{ij}^{pv}\label{t_MCS-1}
\end{equation}
where the sum runs over the 8 possible jumps. This time must be
rescaled to take into account the real vacancy concentration, which
depends on the precipitation microstructure. If one assumes that the
vacancy concentration remains at equilibrium in the different phases
during all the precipitation, a convenient way to perform this time
rescaling is :
\begin{equation}
t=t_{MCS}\frac{c_V^{MC}({\rm Fe})}{c_V^{eq}({\rm
Fe})}\label{t_MCS-2}
\end{equation}
where  $c_V^{eq}({\rm Fe})$ is the equilibrium vacancy concentration
in pure iron and $c_V^{MC}({\rm Fe})$ is the vacancy concentration
in the copper-free regions of the KMC simulation box
\cite{LBS-02,SF-07}. The number  $N_p$ of copper-rich precipitates
and their average size  $\langle i\rangle$ (discussed in Sec. V A)
are computed by considering only clusters which contain  $i\geq p$
copper atoms connected by at least one nearest-neighbor bond.

\subsection{Methods of SSA simulations}

All SSA simulations were performed in the cubic box containing
\,N=$2\times (64)^3$\,  BCC lattice sites with periodic boundary
conditions. In describing precipitates, the precipitate containing
\,$i\geq p$\, copper atoms is defined
 as a set of adjacent sites \,$i$\, (connected by at least one
bond) for which their mean occupations \,$c_i$\, exceed a certain
cut-off value: \,$c_i\geq c_{cut}$,\, while this set contains not
less than \,$p$\, copper atoms:
\begin{equation}
N_{\rm Cu}=\sum_{i}c_i \geq p. \label{ppt-def}
\end{equation}
The choice of \,$c_{cut}$\, was found to be not essential, and for
definiteness
 we took it the same as in experimental studies \cite{Isheim-06}:
 \,$c_{cut}=0.05$.

In solving the SKE (\ref{SKE}) with the diffusional term (\ref{delta
n^d}), we should take into account that this term is proportional to
product of the generalized mobility   \,$M_{ij}$\, and the factor
 \,2\,$\sinh[\beta(\lambda_i-\lambda_j)/2]\simeq \beta(\lambda_i-\lambda_j)$\, describing
a thermodynamic driving force, while mobilities  \,$M_{ij}$\,  given
by Eqs. (\ref{M_ij-a}) or (\ref{M_ij-b}) are proportional to the
correlator   \,$b^{\alpha}_{ij}$ \,  or  \,$b^h_{ij}$\, very sharply
rising with the local concentrations \,$c_i$;\, the latter is
illustrated by Eq. (\ref{b_ij^p-delta}) and by two last columns of
table III. These very sharp dependences do not allow us to use for
solving the SKE the standard iterative methods, such as the Euler or
Runge-Kutta ones: after several iterations, the product
\,$M_{ij}\beta(\lambda_i-\lambda_j)$\, becomes so large that  the
time step needed to achieve a numerical stability gets too small for
these algorithms can be used. On the other hand, there is no
physical reason for the diffusional term to be too large, as the
high mobility \,$M_{ij}$\, should lead to a very fast approaching
the local equilibrium state at which local chemical potentials of
adjacent sites, $\lambda_i$ and $\lambda_j$, are very close to each
other. Thus, in reality diffusional terms (\ref{delta n^d}) remain
to be reasonably small. The problem with application of standard
iterative methods arises due to their discrete nature, thus
numerical values of differences \,$|\lambda_i-\lambda_j|$\, can not
catch up with a very fast increase of \,$b_{ij}$,\, and the
fictitious increase of their product happens.

To overcome this methodical difficulty,  in our iterative
computations we put restrictions on \,$b_{ij}$\,  setting it to not
exceed a certain value \,$b_{ij}^{max}$:\, \,$b_{ij}\leq
b_{ij}^{max}$.\, Then we made simulations with different
\,$b_{ij}^{max}$,\, from smaller to larger ones, until all
physically important characteristics of evolution, including the
incubation and nucleation time, \,$t_r^{inc}$\,  and \,$t_r^N$,\,
and the maximum density of supercritical precipitates,
\,$d_s^{max}$,\, ceased to significantly change under increase of
\,$b_{ij}^{max}$.\, It happens at  \,$b_{ij}^{max}=250$,\, and this
value was used in all our simulations.\, Actually, at
\,$b_{ij}^{max}=500$,\, the \,$d_s^{max}$\, values may be even lower
than at \,$b_{ij}^{max}=250$\, (while for \,$b_{ij}^{max}<250$,\,
they monotonously increase with \,$b_{ij}^{max}$)\, but differences
lie within statistical errors. After the nucleation stage is over
and fluctuations are switched off according to model
(\ref{g_t-t_r^N}), values $|\beta(\lambda_i-\lambda_j|$ typically
decrease by two orders of magnitude with respect to the nucleation
stage. This allows us to significantly decrease the
\,$b_{ij}^{max}$,\,  usually up to $b_{ij}^{max}=10$, and thus to
increase the time-step in solving equations. Again we checked that
using the same \,$b_{ij}^{max}$\, equal to 250 both before and after
\,$t_r^N$,\, and reducing it at \,$t_r>t_r^{N}$\, up to
\,$b_{ij}^{max}$=10,\, lead to the same description of  evolution.
In our case, using the Runge-Kutta method does not provide any
improvement of  stability for the numerical solution of equations
while it needs more calculations
 at each integration step, so in our simulations we used the
more simple Euler method.

Finally, let us discuss the rescaling of the reduced time  \,$t_r$\,
used in the SSA  to the physical time \,$t$\, determined by Eq.
(\ref{t-t_r}). In accordance with the remarks in the end of Sec. II
C, we suppose the effective mean time of direct atomic exchanges,
\,$\tau_{\alpha h}^{\rm eff}$\, in (\ref{t-t_r}),
 to be constant before and after nucleation (that is, at both
 \,$t_r<t_r^{inc}$\,  and   \,$t_r>t_r^N$),\,
and to linearly vary with \,$t_r$\, in the course of nucleation:
\begin{equation} {\tau_{\alpha h}}=
\left\{
\begin{array}{ll}
a_1, &t_r < t_r^{inc};\\
a_1 + \displaystyle{(a_2 -
a_1)\frac{(t_r-t_r^{inc})}{(t_r^N-t_r^{inc})}}\,,\
\hskip3mm  & t_r^{inc} \leq t_r \leq t_r^N;\\
a_2, & t_r > t_r^N. \\
\end{array}
\right.\label{dt/dt_r}
\end{equation}
The constant \,$a_1$\, is the ratio of physical and reduced
incubation times: \,$a_1=t_{inc}/t_r^{inc}$,\, while the constant
\,$a_2$\, can be estimated from the fit of the SSA simulation
results to the evolution rate at the coarsening stage. Thus, model
(\ref{dt/dt_r}) includes only two parameters, \,$t_{inc}$\,  and
\,$a_2$,\, which can be estimated either from comparison to KMC
simulations or from experiments.

In this work we use for such estimates the KMCA results described
below. The resulting rescaling of time is presented in Fig.
\ref{t_r-t} as  the dependence of effective direct exchange rates
\,$\gamma_{\alpha h}^{\rm eff}$\,
 in Eq. (\ref{t_r}) (to be abbreviated \,$\gamma^{\rm eff}$)\,
on the physical time \,$t$;\, this dependence has a more clear
physical meaning than  \,$\tau_{\alpha h}(t_r)$\, in Eq.
(\ref{dt/dt_r}). The results presented in Fig. \ref{t_r-t} and in
analogous  Fig. 3 of BV \cite{BV-98} clearly illustrate the decisive
role of vacancy trapping
 for temporal dependences of effective direct
exchange rates. For all models considered, these rates monotonously
increase with the evolution time  \,$t$\, which reflects the
development of vacancy trapping in the course of precipitation. For
the BV model, this trapping is relatively weak, thus \,$\gamma^{\rm
eff}$\, increases to the coarsening stage just by about twice. For
the LBS model, the vacancy trapping is also not too pronounced
(though stronger than for the BV model), thus the full increase of
\,$\gamma^{\rm eff}$\,  is about 6 times. For the more realistic SF
model, the vacancy trapping  is very strong
 \cite{SF-07}. Thus for all three SF states
considered, the \,$\gamma^{\rm eff}$\, values rise between the
incubation and nucleation stages by more than two orders of
magnitude, and with lowering  temperatures \,$T$\,  this rise seems
to increase, in accordance with a  probable more strong trapping  at
lower \,$T$.

\begin{figure}
\caption {(color online). Temporal dependence of effective direct
exchange rate \,$\gamma_{\rm CuFe}^{\rm eff}$\, in Eq.
(\ref{gamma_pq^eff}) estimated from comparison to the KMCA results
as described in the text. Different curves from top to bottom
correspond to the alloy states LBS-1, SF-1, SF-2 and SF-3,
respectively. Arrows indicate the incubation time \,$t_{inc}$.\label
{t_r-t}}
\end{figure}

\section{EVOLUTION OF MICROSTRUCTURE OBSERVED IN KMCA AND
 SSA SIMULATIONS}

\subsection{Evolution of density and sizes of precipitates}

\begin{figure}
\caption {(color online). Evolution of density of precipitates
containing \hbox{\,$i\geq p$}\, copper atoms,  \,$d_p$\, (left
scale), and the total number of such precipitates within the KMC
simulation box
 \,$V_s^{\rm KMC}$, \,$N_p=N_p^{\rm KMC}$\,  (right scale), for the
\hbox{SF-1}  alloy state. For the SSA simulations with \,$V_s^{\rm
SSA}$=$2V_s^{\rm KMC}$,\, these \,$N_p$\, should be doubled:
\,$N_p^{\rm SSA}$=2$N_p$.\, Solid lines from top to bottom (green,
red, blue and pirple online) correspond to the KMCA results for
\,$p$\,=\,11, 15, 21 and 26, respectively, while symbols (circles,
squares, triangles and crosses) show the analogous results obtained
in another KMC run. Dashed line shows the SSA results for
\,$p=15$,\,
 \,$g_0=1.65$.\,  \label{d_p-SF1}}
\end{figure}

\begin{figure}
\caption {(color online). Same as in Fig. \ref{d_p-SF1} but for the
\hbox{SF-2} state and \,$p$\,=\,8, 11, 16 and 21, respectively.
Dashed curve
 shows the SSA results for  \,$p=11$,\,  \,$g_0=1.7$,\, and chained line,
 the SSA results for \,$p=11$,\,  \,$g_0=1.8$.\, \label{d_p-SF2}}
\end{figure}

\begin{figure}
\caption {(color online). Same as in Fig. \ref{d_p-SF1} but for the
\hbox{SF-3} state and \,$p$\,=\,6, 8, 11, 16 and 21, respectively,
while dashed curve shows the SSA results for \,$p=8$,\,
\,$g_0=1.9$. \label{d_p-SF3}}
\end{figure}

\begin{figure}
\caption {(color online). Same as in Fig. \ref{d_p-SF1} but for the
LBS-1 state and \,$p$\,=\,6, 11, 15 and 26, respectively, while
dashed curve
 shows the SSA results for  \,$p=15$,\,  \,$g_0=1.9$. \label{d_p-LBS1}}
\end{figure}

Evolution of density of different precipitates is shown in Figs.
\hbox{\ref{d_p-SF1}-\ref{d_p-LBS1}}. Solid lines and symbols in
these figures show the results obtained in two different  KMC runs;
differences between these
 lines and symbols illustrate the scale of
errors (mainly, statistical) of these KMC results. Fig.
\ref{d_p-SF1} shows that for the SF-1 state and relatively small
 precipitates, \,$p\lesssim 10$,\, such errors at the nucleation stage
are  significant. while at larger \,$p$,\,  and at
 later stages of evolution, these errors decrease.
For the \hbox{LBS-1} model (Fig. \ref{d_p-LBS1}), differences
between two KMC runs are  lower as the fluctuation effects here are
much weaker. Let us also note that the difference
 between values of \,$d_{p}$\, at two neighboring curves,
\,$d_{p_1}(t)$\, and   \,$d_{p_2}(t)$\, in Figs.
\hbox{\ref{d_p-SF1}-\ref{d_p-LBS1}}:  \,$d(i)=(d_{p_1}-d_{p_2}),$\,
has the meaning of the density of precipitates that include  \,$i$\,
copper atoms with  \,$i$\, between \,$p_1$\, and \,$p_2$:\,
\,$p_1\leq i<p_2.$\,

Let us first discuss the KMCA results presented in Figs.
\hbox{\ref{d_p-SF1}-\ref{d_p-LBS1}}. First, we note that they
qualitatively agree with the classical  theory nucleation (CTN)
ideas described in Sec. III A. In particular, all dependences
\,$d_p(t)$\, reveal presence of four main stages of decomposition:
incubation, nucleation, growth and coarsening. For more quantitative
comparison to the CTN, we should estimate the ``critical'' embryo
size  \,$p_c$.\, In thermodynamic calculations
\cite{Cahn-Hilliard,DV-98a,DV-98b}, the critical embryo is precisely
defined as the set of mean occupations  \,$\{c_i\}$\, that
corresponds to the saddle-point of the generalized free energy
\,$F\{c_i\}$\, in the multi-dimensional space  \,$c_i$\,
\cite{DV-98a}. At the same time,  for nucleating precipitates the
analogous
 critical size can not be defined uniquely. It seems natural to define it
 as the lowest value of the embryo size \,$p$\,  such that
the most probable evolution path at \,$p>p_c$\, is growth rather
than shrinkage or dissolution, but due to the presence of
fluctuations inherent to the nucleation process, such ``kinetic''
critical sizes can be determined just approximately.

Using KMCA results presented in Figs.
\hbox{\ref{d_p-SF1}-\ref{d_p-LBS1}}, we can estimate these \,$p_c$\,
from the shape of curves  \,$d_p(t)$\, at different \,$p$.\, For the
SF model, for which fluctuations in dependences \,$d_p(t)$\, are
pronounced at small \,$p$\, and decrease at larger \,$p$,\, we can
suggest that beginning of decreasing these fluctuations with
increasing  \,$p$\, should correspond to the relation  \,$p\gtrsim
p_c$.\, For the states \hbox{SF-1}, \hbox{SF-2} and \hbox{SF-3}, it
corresponds to \,$p_c\sim 15$,\, \,$p_c\sim 11$,\, and \,$p_c\sim
8$.\, As these estimates agree well with the ``thermodynamic''
critical sizes \,$N_c$\, in Table II,  we will use for the critical
size
 \,$p_c$\, in the \hbox{SF-1}, \hbox{SF-2} and \hbox{SF-3} alloy
state the value \,15, 11 and 8, respectively. Note that in the SSA
simulations, the  \,$p_c$\, values are used only to define the ``end
of nucleation'' time  \,$t_r^N$\, in Eq. (\ref{g_t-t_r^N}) as the
time of creation of the last critical precipitate, and slight
variations of  \,$p_c$\, have almost no effect on evolution.

Figs.  \ref{d_p-SF1}-\ref{d_p-SF3} show that temporal dependences
 \,$d_{p}(t)$\, at these \,$p$=$p_c$\, have, generally, a common form
similar to that obtained by SM \cite{SM-00} for their simplified
model that agrees with the CTN ideas described in Sec. III A. At the
same time, these dependences reveal at least two differences from
the CTN ideas.

(A) Fluctuations in dependences \,$d_{p_c}(t)$\, are rather
pronounced and significant even for  ``supercritical'' embryos with
\,$p>p_c$,\, and the more so for undercritical ones with
\,$p<p_c$.\, It is particularly clear seen for the KMC results
presented in Fig.  \ref{d_p-SF1} by solid lines, while in Figs.
\ref{d_p-SF2} and \ref{d_p-SF3} these fluctuations are partly
smoothed due to the plotting of lesser number of the results.\,

(B) These fluctuations are large and important not only during
nucleation but also after its completion when the average density of
supercritical precipitates ceases to increase. It is seen, in
particular, in Fig. \ref{d_p-SF3} where curve \,$d_p(t)$\, at
\,$p=16\simeq 2p_c$\, reveals a wide and pronounced minimum between
\,$t$\,=\,$3\cdot 10^5$\, and \,$4\cdot 10^5$\, sec falling here by
about 1.5 times. Therefore, for  the SF model, the fluctuations are
important for evolution even at the
 growth stage.

Let us now discuss dependences \,$d_p(t)$\, for the \hbox{LBS-1}
state for which the thermodynamic nucleation barrier \,$F_c$\,
(given in Table II) is low:  \,$F_c\lesssim T$.\, In this case,
evolution of not large ``thermodynamically supercritical'' embryos
with  \,$p\gtrsim N_c$\, again differs from the CTN ideas:  effects
of fluctuations here are strong due to the low thermodynamic gain
under growth of such embryos. As their stability is low, they often
dissolve rather than steadily grow as the CTN supposes. It can
qualitatively explain a significant decrease of ``plateau'' in
curves \,$d_p(t)$\, with \,$p\geq$\, 15\, as compared to \,$p$\,=6\,
and  \,$p$\,=10\, in Fig.  \ref{d_p-LBS1}. For this case, it seems
more adequate to suppose the kinetic  critical size \,$p_c$\, to
significantly exceed $N_c$,\, so that the dominant evolution path at
\,$p>p_c$\, would be growth of the embryo. For the \hbox{LBS-1}
state, we estimate this  size  as \,$p_c\simeq 15$.\,

The SSA results for the precipitate density  \,$d_{p}(t)$\,
presented in Figs.  \ref{d_p-SF1}-\ref{d_p-LBS1} have been obtained
as described in Sec. IV. Note that rescaling of
 time \,$t_r$\, used in the SSA to the time \,$t$\, used in the KMCA
(illustrated by Fig. \ref{t_r-t}) corresponds to the above-described
two-parametric fit of only ``horizontal'' intervals
 in Figs. \ref{d_p-SF1}-\ref{d_p-LBS1}, while ``vertical''
intervals, that is, the density \,$d_{p}$\, values, are calculated
in the SSA with no adjustable parameters.  We see that for all four
alloy states considered, these \,$d_{p}(t)$\,  agree with the KMCA
results within  errors of these results mentioned above.  Fig.
\ref{d_p-SF2} (as well as frame \ref{i(t)}$b$
 below)  also shows that  changes in the SSA results
due to possible variations of \,$g_0$\, between \,$g_0$=1.7\, and
\,$g_0$=1.8\, mentioned in derivations of estimates
(\ref{g_0-numbers})\, are relatively small.

\begin{figure}
\caption {(color online). Average number of copper atoms within a
precipitate, $\langle i\rangle$, versus the evolution time \,$t$\,
(in seconds). Frames $a$, $b$, $c$ and $d$ correspond to the alloy
states SF-1, SF-2, SF-3 and LBS-1, respectively. Solid lines and
squares correspond to the KMCA. Dashed lines correspond to the SSA
with \,$g_0$\, from (\ref{g_0-numbers}), and chained line (in frame
$b$), to \,$g_0$=1.8.\, Arrows show the values of \,$t$\, that
correspond to \,$t_r^N$\, in Eq. \protect{(\ref{g_t-t_r^N})}.
\label{i(t)}}
\end{figure}

In Fig. \ref{i(t)} we show temporal dependences of average number of
copper atoms within a precipitate (``precipitate size'') \,$\langle
i\rangle$=$i_p(t)$.\, Results of the  KMCA and SSA calculations seem
usually to agree within the KMCA errors except for time intervals
just after \,$t_r^N$\, where \,$i_p^{\rm SSA}$\, notably exceed
\,$i_p^{\rm KMCA}$.\, This disagreement seems to be again related to
crudeness of the oversimplified model (\ref{g_t-t_r^N}) for the
length \,$g(t_r)$\, determining scale of fluctuative terms
 \,$\delta n_f$\, in the stochastic equation  (\ref{SKE}).
As mentioned, model (\ref{g_t-t_r^N}) corresponds to the constant
\,$g(t_r)=g_0$\, at  \,$t_r<t_r^N$\, and to a sharp increase of
\,$g$\, (and thus to practically abrupt switching-off fluctuations)
at \,$t_r\geq t_r^N$,\, while actually the length \,$g$\, starts to
increase (and fluctuations \,$\delta n_f$\, to decrease) immediately
after beginning of nucleation, and \,$g$\, remains to be finite (and
nucleation effects to be noticeable) at the growth stage, too. Thus
at  \,$t_r>t_r^N$,\, the SSA-calculated \,$i_p(t)$\, in Fig.
\ref{i(t)} grow more rapidly than in the KMCA (and in reality) as
actually this growth is still hampered by noticeable fluctuation
effects. Therefore, we can expect that these disagreements will be
reduced or vanished if the  models of \,$g(t_r)$\, more realistic
than (\ref{g_t-t_r^N}) will be used  in the SSA.

\subsection{Features of microstructure at different stages of precipitation}

\begin{figure}
\caption {(color online). Upper frame: Distribution of copper atoms
for the SF-1 alloy state just before nucleation,  \,$t$=55 sec.,
observed in the KMCA. Each sphere (blue online) corresponds to a
copper atom. Lower frame: Distribution of local concentrations
\,$c_i$\, for the same state as in the upper frame observed in the
SSA. Relation between coloring and \,$c_i$\, value on the lattice
site \,$i$\, is shown in the right part of the frame. Sites with
\,$c_i<0.05$\, are not shown. \label{evol-1}}
\end{figure}
\begin{figure}
\caption {(color online). Same as in Fig. \ref{evol-1} but at the
end of nucleation,
 \,$t$=1000 sec. Light spheres (yellow online) in the upper frame
show copper atoms that belong to some supercritical precipitate.
\label{evol-2}}
\end{figure}
\begin{figure}
\caption {(color online). Same as in Fig. \ref{evol-2} but at the
beginning of coarsening,
 \,$t$=2000 sec.  \label{evol-3}}
\end{figure}
\begin{figure}
\caption {(color online). Same as in Fig. \ref{evol-2} but at some
intermediate stage of coarsening,
 \,$t$=4000 sec.  \label{evol-4}}
\end{figure}

In Figs. \ref{evol-1}-\ref{evol-4} we show microstructure
 of the SF-1 alloy state at different stages of evolution
observed in the KMCA and  SSA-based simulations. As the SSA
simulation box was twice as large as that used in the KMCA, the
total number of precipitates in the SSA-based (lower) frame of each
of Figs. \ref{evol-1}-\ref{evol-4} exceeds that presented in the
KMC-based (upper) frame by about twice. Figs.
 \ref{evol-1}-\ref{evol-4} illustrate processes of nucleation,
growth and coarsening discussed above.

In comparison of the KMCA and  SSA results in Figs.
\ref{evol-1}-\ref{evol-4} we note that the SSA, as well as any other
statistical description, disregards details of particular atomic
configurations presented in the KMCA snapshots, but it enables us to
pick out essential features of microstructure which often can not be
easily apprehended in these snapshots. It is illustrated by Fig.
\ref{evol-1} where the SSA frame clearly shows a number of regions
significantly enriched by copper atoms, that is, precursors of
precipitates to be created, while such regions are not clearly seen
in the analogous KMC frame. Similar differences in describing
precipitates can be seen in Figs. \ref{evol-2}-\ref{evol-4}. Main
quantitative characteristics of precipitates, such as their density
and sizes discussed above, can be easily determined in both
approaches. At the same time, the presence of significant
crystalline anisotropy, particularly for not large precipitates, in
the KMCA can be established only after some special analysis
\cite{SBM-96,SF-07}, while in the SSA frames it is seen at first
sight. Difference between two descriptions is also evident for
interfaces of precipitates. In the KMCA frames of Figs.
\ref{evol-2}-\ref{evol-4}, these interfaces are usually rather sharp
but  typically not flat and not regular, while in the analogous SSA
frames, these interfaces seem to be somewhat diffuse and have
typically ``intermediate'' values of local concentration: \,$c_i\sim
0.5$\, (green online). The differences arise because each local
concentration \,$c_i$\, in the SSA is obtained by averaging over
some locally equilibrated vicinity of site \,$i$,\, that is, over a
relatively rapid motion of surface atoms on the ``not-filled'' facet
considered. At the same time, inner parts of precipitates (for both
the KMCA and SSA) include only copper atoms which in the SSA frames
is clearly seen on cuts of precipitates by the boundary planes of
simulation box.

Therefore, Figs. \ref{evol-1}-\ref{evol-4} also illustrate a
complementary character of describing evolution of microstructure by
the KMCA and the SSA.

\subsection{Kinetics of nucleation}

As discussed above, the SSA provides a partly averaged description
of atomic distributions aiming mainly at adequate calculations of
locally averaged quantities, such as the density of different
precipitates, their structure, morphology, etc. Therefore,
quantitative treatments of phenomena which are mainly
 determined by fluctuations, such as processes of creation and
evolution of under-critical and near-critical embryos, lies,
generally, outside the scope of the SSA.  However, as mentioned
above, the qualitative changes of microstructure, such as those
corresponding to the nucleation process, can often be more easily
followed in the SSA rather than in the KMCA. Therefore, it can be
instructive to study kinetic details of nucleation
 with the use of the SSA, even though the scale of fluctuation effects in
such study is most probably underestimated.

\begin{figure}
\caption {(color online). Distribution of local concentrations
 within a plane containing several nucleating precipitates
for the SF-1 state at the following  reduced time \,$t_r$\, values:
(a) 7.\,5, (b) 7.\,9, (c) 8.\,0,  (d) 6.\,1, (e) 8.\,5, and (f)
9.\,0. For each point  \,$\bf r$\, of the figure, the \,$c({\bf
r})$\, value is obtained by interpolation between \,$c_i$\, on
neighboring lattice sites. Relation between coloring and \,$c_i$\,
values is shown in the right part of the figure.\label{nucl-1}}
\end{figure}

Some results of such studies are presented in Figs.
\hbox{\ref{nucl-1}-\ref{nucl-3}} which illustrate processes of
sequent creation of three supercritical embryos for the SF-1 alloy
state. Frames \ref{nucl-1}a-\ref{nucl-1}e also show processes of
creation and dissolution of an ``undercritical'' embryo: the
concentration fluctuation in the right central part of these frames
first increases up to the values \,$c_i\sim 0.1$,\, but then
decreases and disappears. On the contrary, frames
\ref{nucl-1}d-\ref{nucl-1}f illustrate a ``successful'' nucleation
process. The local fluctuation  below that discussed above  first
increases in both size and amplitude, and then suddenly shrinks with
formation of a ``kinetic'' supercritical embryo. Later on this
embryo survives and grows, but this growth is first non-monotonic
and includes a ``partial dissolution'' process illustrated by frames
\ref{nucl-2}a and \ref{nucl-2}b. Note that at first stages of this
process shown in frames \ref{nucl-1}e
 and \ref{nucl-1}f, the embryo is extended and shapeless, while
later on it is rather wrong-shaped,thus it seems to have
 little in common with the ``thermodynamic'' critical
embryos shown in Fig. \ref{crit-embryos}.

Creation and evolution of two other embryos shown in Figs.
\ref{nucl-2} and \ref{nucl-3} proceeds similarly. In both cases, the
initial extended and shapeless fluctuation of concentration first
suddenly shrinks so that maximum concentrations within it reach
``critical'' values \,$c_i\gtrsim 0.12$,\, after which embryos
 seem to become ``supercritical''. Then subsequent fluctuations lead to a partial
dissolution of these embryos, but in both cases they survive and
later on start to grow. This growth correspond to ``sucking'' of
copper atoms from adjacent regions and thus to depletion of copper
concentration in these regions.

The examples considered can also illustrate the kinetic mechanism of
nucleation. It can be viewed as a local spinodal decomposition that
starts when the amplitudes and extension of local fluctuative
enrichment of concentration become large enough in order that the
``uphill diffusion'' mechanism characteristic of spinodal
decomposition \cite{VBD} becomes operative. For the examples
considered, it seems to correspond to local concentrations
\,$c_i\gtrsim$\,0.07-0.09 (while the uniform spinodal decomposition
boundary, according to Fig.\ref{ph-diag-SF}, is
\,$c_s\simeq$\,0.045) in the region \,$l\gtrsim (7-8)\,a$.\, Such
interpretation of nucleation as a fluctuation-induced local spinodal
decomposition can be useful for qualitative understanding of many
phenomena in this field, in particular, of a great excess in
probabilities of nucleation near the binodal curve observed
experimentally  \cite{Miyazaki} with respect to  estimates of the
classical theory of nucleation (\ref{W_c}).

\begin{figure}
\caption {(color online). Same as in Fig. \ref{nucl-1} but at the
following \,$t_r$:\, (a) 21, (b)  24, (c)  29, (d) 31, (e)  34, and
(f) 38. \label{nucl-2}}
\end{figure}

\begin{figure}
\caption {(color online). Same as in Fig. \ref{nucl-1} but at the
following \,$t_r$:\, (a) 46, (b) 47, (c) 48, (d) 49, (e) 50, and (f)
53. \label{nucl-3}}
\end{figure}

\subsection{Changes in microstructural evolution under variations
of temperature or concentration}

As mentioned in Sec. IV A, the kinetic type of alloy decomposition
is mainly determined by the  value of the reduced supersaturation
\,$s$\, (\ref{s}), which for metastable states under consideration
is less than unity. Low  \,$s$\, correspond to the states with the
concentration and temperature  \,$(c,T)$\, values close to the
binodal curve for which a ``deep NG'' type of evolution with
 a low density and large sizes of nucleating precipitates
 is characteristic. An increase of \,$s$\, corresponds to
approaching the spinodal curve  and decreasing  nucleation barriers,
thus nucleating precipitates should become smaller (which is
illustrated by Table II), while their density should increase.
Basing on these considerations, we can expect that the reduced
supersaturation \,$s$\, is the main parameter determining
microstructural evolution, but at the given  \,$s$,\, microstructure
can also significantly vary with the concentration or temperature.

\begin{figure}
\caption {(color online). Distribution of concentrations \,$c_i$\,
at the end of nucleation for the following alloy states: (a) SF-1,
(b) SF-3, and (c) SF-4. \label{evol-c,T}}
\end{figure}

To get an idea about these varuations,  in Fig. \ref{evol-c,T} we
show microstructure at the end of nucleation for the SF-1, SF-3 and
SF-4 alloy states described by Tables I, II and Fig.
\ref{crit-embryos}. In simulations for the SF-4 state (which has has
the same supersaturation as the SF-3 state but higher concentration
and temperature) we used \,$g_0=1.9$\, value, same as for the SF-3
state. In accordance with considerations mentioned above, the
increase of supersaturation  \,$s$\,  in states SF-3 and SF-4 with
respect to SF-1 by about 1.5 times leads to much higher density of
nucleated precipitates: by about 3.6 times for the SF-3 state, and
by 2.2 times for the SF-4 state. At the same time, frames
\ref{evol-c,T}b and \ref{evol-c,T}c  illustrate differences in
microstructure for
 the same \,$s$\, but different \,$c$\,  and \,$T$.\, For the state SF-4,
precipitates are notably larger while their density is by 1.5 times
lower than those for the state SF-3. These differences can be
qualitatively explained by the differences in characteristics of
thermodynamic critical embryos for these two states presented in
Table II and Fig. \ref{crit-embryos}: both  critical sizes  and
reduced nucleation barrier \,$F_c/T$\, and for the SF-4 state
notably exceed those for the SF-3 state.

\section{CONCLUSIONS}

To conclude, we summarize the main results of this work.

1. The consistent and computationally efficient stochastic
statistical approach is developed to microscopically study kinetics
of decomposition of metastable alloys.

2. In this approach, description of evolution in terms of certain
reduced time includes no adjustable parameters. Rescaling of this
reduced time to the physical time can usually be made with the use
of few constants which can be estimated either from comparison to
kinetic Monte Carlo simulations or from experiments.

3. For several realistic models of iron-copper alloys studied, the
results of this approach usually agree with the kinetic  Monte Carlo
simulation results within errors of these simulations.

4. Oversimplified model (\ref{g_t-t_r^N}) for the important kinetic
parameter of the theory, size of locally equilibrated
 regions, seems to be sufficient for describing main
characteristics of microstructure. However, for an adequate
description of temporal dependences we should use more realistic
models discussed in Secs. IV B and V A.

\
\section*{ACKNOWLEDGMENTS}

The authors are much indebted to I.A. Zhuravlev for his help in this
work, as well as to Yu. N. Gornostyrev,  P. A. Korzhavy and Georges
Martin, for numerous valuable discussions. The work was supported by
the Research Technological Center ``Ausferr'', Magnitogorsk, by the
Russian Fund of Basic Research (grant No. 06-02-16476); by the fund
for support of leading scientific schools of Russia  (grant No.
NS-3004.2008.2); and by the program of Russian university scientific
potential development (grant  No. 2.1.1/4540).

\end{document}